\documentclass[twocolumn, a4size, 10pt]{IEEEtran}	

\usepackage{fancyhdr, epsfig, epsf, amsthm, amsmath, amssymb, amsfonts, color}
\usepackage{algpseudocode,algorithm,algorithmicx}
\usepackage{threeparttable}
\usepackage[noadjust]{cite}
\usepackage{dsfont}
\usepackage{enumerate}
\usepackage{comment}
\usepackage{multirow}
\usepackage{bigstrut}
\usepackage{tcolorbox}
\usepackage{graphicx,lscape,rotating}
\usepackage[caption=false]{subfig}

\newtheorem{proposition}{Proposition}
\newtheorem{remark}{Remark}

\newtheorem{assumption}{Assumption}

\def\qed{$\blacksquare$}

\def\endproof{\hfill \qed}

\def\E{\mathsf{E}}
\def\SINR{\mathsf{SINR}}
\def\SNR{\mathsf{SNR}}

\def\l{\left}
\def\r{\right}
\def\({\left(}
\def\){\right)}

\def\[{\left[}
\def\]{\right]}

\def\ba{{\mathbf{a}}}
\def\bb{{\mathbf{b}}}

\def\bee{{\mathbf{e}}}

\def\bh{{\mathbf{h}}}

\def\bq{{\mathbf{q}}}

\def\bs{{\mathbf{s}}}

\def\bv{{\mathbf{v}}}

\def\bx{{\mathbf{x}}}

\def\bz{{\mathbf{z}}}
\def\b0{{\mathbf{0}}}

\def\bB{{\mathbf{B}}}
\def\bC{{\mathbf{C}}}

\def\bE{{\mathbf{E}}}
\def\bF{{\mathbf{F}}}

\def\bH{{\mathbf{H}}}
\def\bI{{\mathbf{I}}}
\def\bJ{{\mathbf{J}}}

\def\bP{{\mathbf{P}}}
\def\bQ{{\mathbf{Q}}}
\def\bR{{\mathbf{R}}}

\def\bT{{\mathbf{T}}}

\def\bX{{\mathbf{X}}}

\def\mC{{\mathbb{C}}}

\newcommand{\nn}{\nonumber}

\def\T{\text{T}}
\def\H{\text{H}}
\def\bb{\text{BB}}
\def\rf{\text{RF}}
\def\f{\mathbf{f}}
\def\tr{\text{Tr}}
\def\tot{\text{tot}}
\def\cf{C_\text{F}}

\def\papertitle{Hybrid Precoding for Massive MIMO Systems in \\Cloud RAN Architecture with Capacity-Limited Fronthauls}

\IEEEoverridecommandlockouts

\begin{document}
\title{ \fontsize{20}{24}\selectfont  \papertitle}

\author{Jihong~Park, Dong~Min~Kim, Elisabeth~De~Carvalho, and Carles~Navarro~Manch\'on
\thanks{J.~Park, D.~Kim, E. Carvalho, and C. Manch\'on are with Department of Electronic Systems, Aalborg University, Denmark (email: \{jihong, dmk, edc, cnm\}@es.aau.dk).
}
}

\maketitle \thispagestyle{empty}

\begin{abstract}
Cloud RAN (C-RAN) is a promising enabler for distributed massive MIMO systems, yet is vulnerable to its fronthaul congestion. To cope with the limited fronthaul capacity, this paper proposes a hybrid analog-digital precoding design that adaptively adjusts fronthaul compression levels and the number of active radio-frequency (RF) chains out of the entire RF chains in a downlink distributed massive MIMO system based on C-RAN architecture. Following this structure, we propose an analog beamformer design in pursuit of maximizing multi-user sum average data rate (sum-rate). Each element of the analog beamformer is constructed based on a weighted sum of spatial channel covariance matrices, while the size of the analog beamformer, i.e. the number of active RF chains, is optimized so as to maximize the large-scale approximated sum-rate. With these analog beamformer and RF chain activation, a regularized zero-forcing (RZF) digital beamformer is jointly optimized based on the instantaneous effective channel information observed through the given analog beamformer. The effectiveness of the proposed hybrid precoding algorithm is validated by simulation, and its design criterion is clarified by analysis.
\end{abstract}

\section{Introduction}

In Cloud RAN (C-RAN) architecture \cite{CM2015,peng2016recent}, mobile devices are connected to one or several remote radio heads (RRHs) which serve as access points to the network.
The RRHs are plugged to a network of wired fronthaul links that connect them to the cloud, meaning a baseband unit (BBU) that centralizes the major part of the processing.
Following the original C-RAN concept~\cite{CM2015}, the RRHs are cheap and easy-to-deploy, performing only basic functions, such as beamforming in a radio-frequency (RF) domain, while the BBU manages all the digital functions, including channel estimation and beamforming in a digital domain.

Massive MIMO \cite{marzetta2010noncooperative,rusek2013scaling} base stations are often envisioned as stand-alone entities. Indeed, as they have a very large number of antennas that potentially create huge spatial degrees of freedom,
they are able to both serve user equipments (UEs) efficiently and manage inter-cell interference in many cellular scenarios.
However, in scenarios with dense device population, a cloud-based distributed architecture is desirable where each RRH is equipped with a large number of antennas. For example, in a megacity, one can imagine compact panels of antennas are deployed along the external walls of buildings,  which are connected to the cloud.
Very large aperture massive arrays can be deployed along the walls of a very large number of infrastructures or around the roof of a stadium, where the very large array would be  made out of smaller panels all connected to the cloud.

In C-RAN architecture, the volume of traffic to be transported via the fronthaul links can be a severe limitation due to the data rate that can be supported by the fronthaul links.
With massive MIMO RRHs, this limitation becomes extremely severe especially when the signals from all antennas should be transported.
It is recognized by \cite{park2017massive,chen2017scalable,KimOsvaldo:17} that a  level of data compression through multiple antenna processing is desirable allowing for a partially centralized solution.
In this paper, we propose a partially centralized solution that
relies on a split of the functionalities in the massive array processing between the RRHs and the BBU based on hybrid analog and digital beamforming \cite{Molish16,Heath14}. Analog beamforming is kept within RRHs, while digital beamforming is
migrated to the BBU. Analog beamforming at each RRH allows a reduction in the number of streams to be forwarded to the BBU.

\begin{figure*}
    \centering
    \includegraphics[width= 18cm]{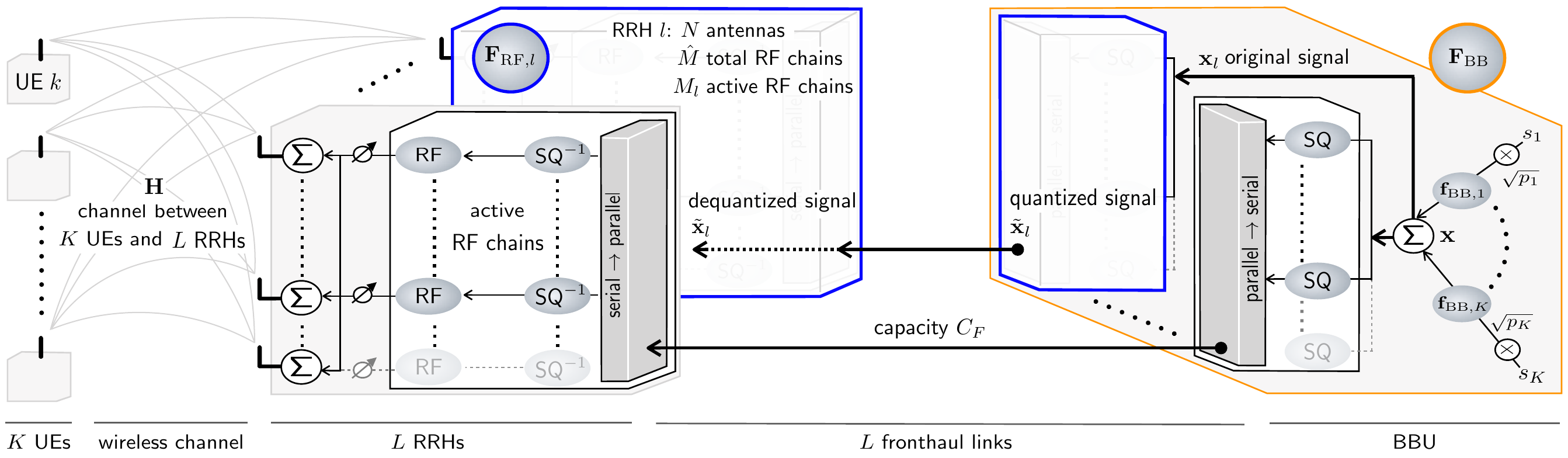}
    \caption{Downlink hybrid precoding C-RAN architecture comprising $K$ UEs jointly served by $L$ RRHs connected to a single BBU through $L$ fronthaul links with capacity $\cf$ per link. At the BBU, different UE messages are linearly combined by applying RZF ($\bF_\bb$) at the BBU. The $l$-th original message signal vector $\bx_l$ is compressed via uniform scalar quantization ($\textsf{SQ}$) up to $M_l$ signals corresponding to the number of active RF chains at RRH $l$. The compressed signals are forwarded to RRH $l$ via a fronthaul link with capacity $\cf$. The forwarded signal is dequantized ($\textsf{SQ}^{-1}$) at RRH $l$ as $\tilde{\bx}$ including quantization noise due to compression. The RRH $l$ applies analog beamforming $\bF_{\rf,l}$, and transmits the signal to $K$ UEs.}
    \label{F:Network}
\end{figure*}

The  goal of our design is to make the BBU determine:
1) the analog beams applied at each RRH,
2) the number of beams that are exchanged between RRHs and the BBU, and
3) the digital beams at the BBU.
The optimization problem is constrained by the transmit power at each RRH as well as the maximal data rate that can be transported through the fronthaul links per RRH. Our design is based on the following principle: the number of beams and  analog beamforming matrices are based on second-order channel statistics, while the digital beams are determined based on the concatenation of the instantaneous channel and analog beams, justified below.

In the usual case where both analog and digital processing parts are colocated, one widely used approach to determine a hybrid analog-digital structure relies on the instantaneous value of channel. As the channel is observed through the analog beamformers, it takes several sequential test measurements to be able to estimate the whole channel \cite{Wang:15,alkhateeb2014channel}. This entails latency with the risk of outdated channel state information (CSI). This problem is aggravated in a C-RAN structure where the training signals are transported to the BBU.
To cope with this problem
we seek an analog beamformer design based only on long-term
CSI, i.e. spatial covariance matrices which are expected
to be accurate for a longer time-frame than instantaneous
CSI. This approach thereby significantly reduces channel
estimation complexity as well as fronthaul consumption for
exchanging CSI. Once the RRHs have adjusted their analog setting, the equivalent channel at the output of the analog beams is estimated at the BBU and forms the basis for the digital beamforming.

The main feature of our hybrid precoding design is its extensively utilizing second-order channel statistics. With this end, we first  formulate an optimization problem  based on the instantaneous value of the channel. As we want the analog design to be based on spatial covariance matrices, we need to modify the objective function.
For that purpose, we assume that  digital beamforming comes from a regularized zero-forcing (RZF) design that can be expressed as a function of the channel and the analog beams. Substituting their expressions in the sum average data rate (sum-rate), we can determine a large-scale approximated sum-rate that only depends on the spatial covariance matrices of the channel for each UE.
This expression is the basis for estimation of both the number of beams and the analog beamforming matrices.

\subsection{Related Works}

Hybrid precoder design has been investigated in
\cite{Alkhateeb2013,Heath:15,ParkHeath:17,Molish16,Tadilo:16,SohrabiWei:15}. Its key
structure is well summarized in \cite{Molish16}, where digital beamforming is connected to RF-domain analog beamformers retaining smaller number of RF
chains compared to a full digital beamformer. The increase in RF chains therefore
improves the performance of a hybrid precoder until reaching its upper bound performance
achieved by a full digital beamformer \cite{Tadilo:16}. In massive MIMO systems with
a general digital beamformer, the minimum number of RF chains to achieve this upper
bound performance has been specified in \cite{SohrabiWei:15} as twice the number of
data streams between digital and analog beamformers. For an RZF digital beamformer,
it has been shown by \cite{ParkHeath:17} that allowing more RF chains is still
beneficial. In massive MIMO C-RAN network, these performance gains induced by allowing more RF
chains may diminish due to the capacity-limited fronthauls connecting digital and
analog beamformers.

In C-RAN architecture, different configuration of precoding function splits between RRHs
and the BBU have been proposed and summarized in
\cite{park2014fronthaul,peng2015fronthaul}, including the design with the RRHs
being capable only of analog beamforming which is of our interest. For a given
precoding design, fronthaul compression schemes to comply with the limited fronthaul capacity have been
investigated via an information theoretic approach
\cite{KimOsvaldo:17,KangSimeone:17} and under the use of a scalar quantizer
\cite{LiuZhang:15,LiuZhangJoint:15}. In these works, fronthaul forwarding
information is compressed and the level of compression is adjusted so as to meet the
fronthaul capacity. More compression, i.e. coarse quantization levels, induces
larger quantization noise, ending up with deteriorating the useful received signal. The
precoder is therefore optimized accordingly, which poses another challenge under
limited fronthaul capacity that may not allow frequent CSI exchange between RRHs and
the BBU and/or bring about outdated CSI.

One promising approach to resolve this precoding design problem under limited fronthaul capacity is to utilize spatial covariance of the channel instead of instantaneous CSI, which has been investigated in \cite{CaireJSDM:13,ParkHeath:17}. Spatial covariance matrix is less frequently changing than instantaneous CSI, and can thus be estimated more easily as shown by \cite{GesbertCovariance:13,Alkhateeb2013,LeeCovariance:14,GuillaudSlock:15,EliasiRappaport:17}. The precoding design based on spatial covariance matrix becomes more effective in massive MIMO systems in C-RAN architecture where their large number of antennas lead to huge amount of CSI to be estimated. As the number of antennas increases, it has been shown by \cite{Wagner:12,HoydisDebbah:13} that the instantaneous signal-to-interference-plus-noise ratio ($\SINR$) asymptotically converges to a deterministic value that is only a function of spatial covariance matrices. Such a deterministic equivalent can be regarded as the large-scale approximation and be exploited for precoding design that is no longer depending on instantaneous CSI \cite{KerretGesbert:15,TabikhSlock:16,TabikhSlock:17}.

A notably relevant work to this paper is \cite{ParkHeath:17} that proposes a hybrid precoding design exploiting spatial covariance matrices. Another remarkable work is \cite{KimOsvaldo:17} that suggests a hybrid precoding design based on instantaneous perfect CSI, while concerning the fronthaul compression effect under limited fronthaul capacity. Bridging the gap between them, in this paper, we consider a distributed massive MIMO system enabled by C-RAN with limited fronthaul capacity, and propose its hybrid precoding design based on spatial covariance matrices. The proposing hybrid precoding design, in distinction from the preceding works, additionally optimizes the RF chain activations to mitigate fronthaul congestion.

\subsection{Contributions and Organization}

Our general contribution is a design of distributed architecture for massive MIMO where the data volume on the fronthaul is managed by relying on a split of the hybrid analog-digital beamforming functions, as well as a design of the analog beams based on second-order channel statistics.
For tractability, the optimization solution is made simple: the analog beams are determined in a disjoint optimization problem, i.e. per-RRH optimization, and the number of analog beams is the only parameter accounting for the fronthaul constraint.

Our major contributions are summarized as follows.
\begin{itemize}

\item
In a multi-user hybrid analog digital setup, where the analog and digital functions are colocated in a signal device, we assume that
 digital beamforming is based on a zero-forcing (ZF) design and derive an optimization criterion for the analog beams based on the spatial covariance of the UEs.
To obtain a tractable solution, we make the link with the method recently proposed in \cite{ParkHeath:17} where the analog beams are determined as the strongest singular vectors of the sum of the spatial covariance matrices of all UEs. Compared to  \cite{ParkHeath:17}, we base the analog beams on a weighted sum of the covariance matrices allowing to account for UE channels with different energy.

\item
For a distributed massive RRHs, we derive the large-scale approximation of the $\SINR$ corresponding to a digital beamforming design based on an RZF, leading to an approximation of the sum-rate.

\item
We optimize the number of analog beams that can be transported through the fronthaul and quantization precision based on the large-scale approximated sum-rate.

\end{itemize}

The paper is organized as follows. In Section~II, we present the system model for a
downlink massive MIMO system based on C-RAN architecture with limited-capacity fronthaul links. In Section~III, we describe the sum-rate maximization problem in this system, and provide its deterministic approximated problem by deriving the large-scale $\SINR$ approximation. \;\;\;\;\;\;In Section~IV, we propose a novel analog beamforming design based on a weighted sum of spatial covariance matrices. \;\;\;\;\;\;\;\;In Section~V, we propose a hybrid precoding algorithm based on spatial covariance matrices and the large-scale approximated $\SINR$. Numerical results are provided in Section~VI, followed by concluding remarks in Section~VII.

\section{System Model}
As Fig.~\ref{F:Network} illustrates, we consider a downlink C-RAN where a single BBU
is connected to $L$ RRHs jointly serving $K$ single-antenna UEs. The RRHs
communicate with the BBU via error-free fronthaul links with identical
capacity $\cf$ bps. Each RRH is equipped with $N>K$ antennas and phase shifters,
which are controlled by up to $\hat{M}\leq N$ RF chains. In the $l$-th RRH for $l\in\{1,2,\cdots, L\}$, only $M_l\leq \hat{M}$ RF chains become active while the
rest are left inactive to minimize the fronthaul usage. In the following
subsections, we specify the channel and hybrid precoding models of the downlink
C-RAN.

For convenience, we define the following notations. The aggregate numbers of antennas and active RF chains of all RRHs are $\bar{M} = \sum_{l=1}^L M_l$ and $\bar{N}=L\cdot N$ respectively. The superscripts $\T$ and $\H$ indicate matrix transpose and conjugate-transpose operations. The notation $[\bX]_{ij}$ indicates the $(i,j)$-th entry of a matrix $\bX$. The matrices $\bI_A\in\mC^{A\times A}$ and $\b0_{A\times B}\in\mC^{A\times B}$ respectively are an identity matrix and zero matrix for non-negative integers $A, B\geq 1$. Other important notations are summarized in Table~I.

\begin{table}
\centering
\caption{List of notations.}
\renewcommand{\arraystretch}{1.3}
\begin{tabular}{r |l }
\bf{Notation} & \textbf{Meaning}\\
\hline
    $K$ & \# single-antenna UEs\\
    $L$ & \# RRHs\\
    $N$& \# antennas per RRH\\
    $\bar{N}$& \# aggregate antennas of all RRHs, i.e. $\bar{N}=LN$\\
    $\hat{M}$& \# RF chains per RRH\\
    $M_l$& \# active RF chains per RRH, $M_l\leq\hat{M}$\\
    $\bar{M}$& \# aggregate active RF chains of all RRHs\\
    $\cf$& Fronthaul capacity per RRH\\
    $P_{\tot}$& Maximum transmit power per RRH\\
    $\bh_{k,l}$&  Channel vector between UE $k$ and RRH $l$\\
    $\bh_k$& Concatenated channel vector between UE $k$ and $L$ RRHs\\
    $\bH$ & Aggregate channel matrix between K UEs and $L$ RRHs\\
    $\bR_{k,l}$& Spatial covariance matrix of UE $k$ at RRH $l$\\
    $\bR_k$& Aggregate spatial covariance matrix of UE $k$ for $L$ RRHs\\
    $\bF_{\rf,l}$& Analog beamforming matrix of RRH $l$\\
    $\bF_{\rf}$& Aggregate analog beamforming matrix of $L$ RRHs\\
    $\f_{\bb,k}$& Digital beamforming vector for UE $k$\\
    $\bF_{\bb}$& Aggregate digital beamforming matrix for $L$ UEs\\
    $\bQ_l$& Quantization noise covariance matrix of RRH $l$\\
    $\bQ$& Aggregate quantization noise covariance matrix of $L$ RRHs\\
    $\bx$& Original signal vector before fronthaul compression\\
    $\tilde{\bx}$& Transmit signal vector after fronthaul compression\\
\end{tabular}\label{Table:Notations}
\end{table}

\subsection{Channel Model}
In the downlink C-RAN, $K$ UEs and a total of $\bar{N}=LN$ antennas construct a
MISO broadcast channel matrix $\bH = [\bh_1,\cdots,\bh_K ]^\H \in\mC^{K\times
\bar{N}}$. Each column vector $\bh_k = [\bh_{k,1}^\T,\cdots,\bh_{k,L}^\T
]^\T\in\mC^{\bar{N}}$ for $k\in\{1,2,\dots,K\}$ denotes the channel vector from all
antennas of all RRHs to UE $k$. The channel vectors $\bh_{k,l}\in\mC^{N}$ between UE
$k$ and RRH $l$ for $l\in\{1,2,\dots,L\}$ are modeled as
\begin{align}\label{Eq:hklDef}
\bh_{k,l} = \sqrt{N } \bR_{k,l}^\frac{1}{2}\bz_{k,l}
\end{align}
where $\bz_{k,l}\in \mC^{N}$ is a  vector with independent and identically distributed (i.i.d.) circularly-symmetric complex Gaussian entries of zero mean and variance $1/N$. The term $\bR_{k,l}=\E[\bh_{k,l} \bh_{k,l}^\H] \in \mC^{N\times N}$ is the spatial covariance matrix between UE $k$ and RRH $l$, capturing large-scale and small-scale fading components in a way that
\begin{align}\label{Eq:RklDef}
 \bR_{k,l} = \delta_{k,l}\mathbf{\Theta}_{k,l}
\end{align}
where $\delta_{k,l} = {d_{k,l}}^{-\eta}$ denotes a path-loss constant between UE $k$
and RRH $l$, with  $d_{k,l}$ being the distance between them and  $\eta\geq 2$
indicating the path-loss exponent, and $\mathbf{\Theta}_{k,l}$ is a fast-fading covariance matrix.

We assume the vectors $\bh_{k,l}$ $\forall k,l$ to be mutually uncorrelated. In this case, the channel vectors between all the RRHs and UE $k$ can be represented as
\begin{align}
\bh_{k} = [\bh_{k,1}^\T, \cdots, \bh_{k,L}^\T ]^\T = \sqrt{\bar{N}} \bR_k^\frac{1}{2}  \bz_{k}
\end{align}
where $\bz_k\in\mC^{\bar{N}}$ is a vector with independent and circularly-symmetric complex Gaussian entries with zero mean and variance $1/\bar{N}$, and $\bR_k\in\mC^{\bar{N}\times \bar{N}}$ is the UE $k$'s aggregate spatial covariance matrix to be seen by the BBU, given as a block diagonal matrix containing $\bR_{k,l}$ as the $l$-th diagonal entry, e.g. $\bR_{k} =\[\begin{smallmatrix} \bR_{k,1} & \b0_{N\times N}\\\b0_{N\times N} & \bR_{k,2} \end{smallmatrix}\]$ for $L=2$.

\subsection{Hybrid Precoding}\label{Sect:Hybrid}
For analog beamforming, we assume that the BBU perfectly knows the long-term global CSI, i.e. having perfect knowledge of the spatial covariance matrices $\bR_k$'s for $k=1,\cdots,K$. As often as the long-term channel statistics change, the BBU determines the RRH $l$'s RF chain activations and analog beamformer $\bF_{\rf,l}\in\mC^{N\times M_l}$ based on $\bR_k$'s, updated via the fronthaul link with a negligible delay. The decision of $\bF_{\rf,l}$ is to be elaborated in Section~IV after clarifying our proposed analog beamforming method based on spatial covariance matrices. For given $\bF_{\rf,l}$'s, we consider the aggregate analog beamformer $\bF_{\rf}\in\mC^{\bar{N}\times \bar{M}}$ seen by the BBU, which is given as a block matrix with all zero entries except for the $L$ diagonal block matrices each of which consists of $\bF_{\rf,l}$, e.g. $\bF_{\rf} =\[\begin{smallmatrix} \bF_{\rf,1} & \b0_{N\times M_2}\\\b0_{N\times M_1} & \bF_{\rf,2} \end{smallmatrix}\]$ for $L=2$.

Next, we assume that the BBU has perfect knowledge of the effective instantaneous channel after analog beamforming, i.e. $\bH \bF_\rf$. Based on this effective instantaneous CSI, the BBU generates the RZF digital beamformer given as
\begin{align}\label{Eq:FbbDef}
\bF_\bb &= \alpha \Big[ \underbrace{(\bH \bF_\rf)^\H \bH \bF_\rf + \bar{N}\beta \bI_{\bar{M}}}_{:=\bC} \Big]^{-1}(\bH \bF_\rf)^\H \\
&= \alpha \bC^{-1} (\bH \bF_\rf)^\H \in \mC^{\bar{M}\times K}.
\end{align}

The term $\alpha>0$ is the amplitude scaling parameter of $\bF_\bb$ assumed as a single constant value for brevity. It is noted that this can be set as a constant diagonal matrix of which each diagonal entity adjusts the amplitude per RRH as in \cite{Heath:15,TabikhSlock:16,TabikhSlock:17}, however deferred to future work. The term $\beta>0$ is the regularization parameter of $\bF_\bb$. This can also be a diagonal matrix made from constant blocks, where each block corresponds to a single RRH. As each RRH sees identical $K$ UEs in the channel model, all the constants for different blocks become identical, leading to this constant $\beta$. Following from \cite{PeelSwindlehurst:15}, $\beta$ is assumed as $K/(\bar{N}\rho)$ for $\rho>0$ being the maximum per-RRH signal-to-noise ratio ($\SNR$), unless otherwise specified. This term can also be optimized as in \cite{Wagner:12}, left the analysis to future work.

The UE $k$'s corresponding digital beamformer is given as
\begin{align}
\f_{\bb,k} = \alpha \bC^{-1}\bF_\rf^\H \bh_k\in \mC^{\bar{M}}. \label{Eq:fbbk}
\end{align}
For the RRH $l$, the effective digital beamforming matrix is
 \begin{align}\label{Eq:FbblDef}
 \bF_{\bb,l}=[\f_{\bb,1,l},\cdots,\f_{\bb,K,l}]^\T  \in\mC^{M_l\times K}
 \end{align}
where $\f_{\bb,k,l}=[f_{\bb,k,l,1},\cdots,f_{\bb,k,l,M_l} ]^\T
~\in~\mC^{M_l}$.

\subsection{Downlink Transmitted Signal after Fronthaul Compression} \label{Sect:SysTransmit}
To enable downlink data transmissions under limited fronthaul capacity, the original
message signal is compressed at the BBU and then forwarded to the RRHs via fronthaul
links. Such fronthaul compression in return incurs additional quantization noise
despite the error-free fronthaul links. The impact of this quantization noise is
specified under a uniform scalar quantization (USQ) model \cite{LiuZhang:15}. The
entire downlink transmission procedure illustrated in Fig.~\ref{F:Network} is
described as follows.

\subsubsection{Digital Precoding}
The BBU allocates transmission powers $\bP = \text{diag}(\sqrt{p_1}, \cdots, \sqrt{p_K})$ to the message symbols $\bs=[s_1, \cdots, s_K]$ that independently follow circularly-symmetric Gaussian distributions with zero mean and unit variance. These messages are linearly combined by applying the RZF precoder $\bF_{\text{BB}}$, resulting in the original message signal vector for all RRHs, which is given as
\begin{align}
\bx =\bF_\bb \bP^\frac{1}{2}\bs = [\bx_1^\T, \cdots, \bx_L^\T]^\T \in \mC^{\bar{M}}
\end{align}
where $\bx_l  = [x_{l,1},\cdots,x_{l,M}]^\T~\in~\mC^{M_l}$ is the original message signal vector to be forwarded to RRH $l$. The $m$-th entry of $\bx_l$ comprises its in-phase (I) and quadrature (Q) components, i.e. $x_{l,m}= x_{l,m}^I + jx_{l,m}^Q$, each of which independently follows a Gaussian distribution with zero mean and variance $ \sum_{k=1}^K p_k |f_{\bb,k,l,m}|^2/2$.

\subsubsection{Fronthaul Compression}
The BBU applies USQ to $\bx_l$ of RRH $l$, which independently quantizes $x_{l,m}^I$ and $x_{l,m}^Q$ by respectively using $D_l$ bits, i.e. $2^{D_l}$ discrete uniform levels. Following \cite{LiuZhang:15}, we consider this quantization process brings about an additional independent Gaussian noise before transferring $\bx_l$, yielding its distorted signal vector given as
\begin{align}
\tilde\bx_l = \bx_l + \bq_l.
\end{align}
The quantization noise vector $\bq_l= [q_{l,1}, \cdots, q_{l,M}]^\H$ follows an independent and circularly-symmetric Gaussian distribution with zero mean and covariance matrix $\bQ_l$, given as
\begin{align}\label{Eq:Ql}
\bQ_l=\E[\bq_l \bq_l^\H] = \text{diag}(\tau_{l,1}^2,\cdots, \tau_{l,M}^2)\in \mC^{\bar{M}\times \bar{M}}
\end{align}
where $\tau_{l,m}^2$ is provided by \cite{LiuZhang:15} as:
\begin{align}
\tau_{l,m}^2 = 3\cdot 2^{-2 D_l}\sum_{k=1}^K p_k |f_{\bb,k,l,m}|^2. \label{Eq:tau}
\end{align} The aggregate distorted signal vector for all the RRHs is given as $\tilde\bx =\bx + \bq \in \mC^{\bar{M}\times 1}$, where $\bx=[\bx_1^\T,\cdots,\bx_L^\T]^\T \in \mC^{\bar{M}\times 1}$, $\bq=[\bq_1^\T,\cdots,\bq_L^\T]^\T \in \mC^{\bar{M}\times 1}$, and aggregate quantization covariance matrix $\bQ=\E[\bq \bq^\H]\in\mC^{\bar{M}\times \bar{M}}$.

\subsubsection{Fronthaul Forwarding and Signal Transmission}
The RRH $l$ has $M_l$ number of active RF chains. The corresponding $M_l$ number of
independent message signal streams originating from the aggregate I-Q components of
$\tilde{\bx}_l$ are parallel-to-serial converted at the BBU. This results in the
single signal stream traffic $2D_l M_l$ bps. In order not to make the generated
traffic exceed the fronthaul link capacity $\cf$, the fronthaul quantization bits
$D_l$ are adjusted accordingly by the BBU. While thereby guaranteeing $2D_l M_l \leq
\cf$, the combined signal stream is forwarded from the BBU to RRH $l$ via the
fronthaul link. At RRH $l$, the combined signal stream is serial-to-parallel
converted in order to be fit with each of the $M_l$ active RF chains, and then is
independently and perfectly restored, resulting in RRH $l$'s transmit signal
vector $\tilde{\bx}_l$.

\subsection{Downlink Received Signal} \label{Sect:SINR}
The RRH $l$ applies its analog beamformer $\bF_{\rf,l}$ to the transmit signal vector $\tilde{\bx}_l$. At UE $k$, the aggregate transmit signals $\tilde{\bx}$ from all the RRHs are experiencing the concatenated channel $\bh_k$, and its received signal is given as

\vspace{-15pt}\small\begin{align}
&y_k = \bh_k^\H \bF_\rf \tilde{\bx} + n_k = \overbrace{\bh_k^\H\bF_\rf\f_{\bb,k} \sqrt{p_k} s_k}^{\text{desired signal}}  \nn\\
&\quad\quad\quad + \underbrace{\bh_k^\H \bF_\rf \sum_{i=1,i\neq k}^K
\f_{\bb,i}\sqrt{p_i}s_i}_{\text{interference}} + \underbrace{\bh_k^\H\bF_\rf \bq
}_{\text{quantization noise}} + \underbrace{n_k}_{\text{Rx noise}}
\end{align}\normalsize
where $n_k$ is thermal noise at reception, following a Gaussian distribution with zero mean and variance~$\sigma^2$. The received $\SINR$ of UE $k$ ($\SINR_k$) is represented accordingly as

\vspace{-10pt}\small\begin{align}
\SINR_k &= \frac{p_k | \bh_k^\H \bF_\rf \f_{\bb,k} |^2}{ \sum\limits_{i=1,i\neq k}^K p_i |\bh_k^\H \bF_\rf \f_{\bb,i} |^2 + \bh_k^\H \bF_\rf \bQ \bF_\rf^\H\bh_k + \sigma^2 }. \label{Eq:SINR_0}
\end{align}\normalsize

\section{Sum-Rate Maximization with Large-Scale $\SINR$ Approximation} \label{Sect:Problem}
In this section, we aim at providing a tractable problem formulation that maximizes the sum-rate of a downlink massive MIMO in C-RAN architecture. With this end, we first formulate the original problem based on the instantaneous $\SINR$, and then approximate the problem by exploiting the large-scale approximated $\SINR$.

\subsection{Original Sum-Rate Maximization with Constraints}
We consider the original sum-rate maximization problem based on the instantaneous $\SINR_k$ provided by \eqref{Eq:SINR_0} in Section~\ref{Sect:SINR} as follows.
\begin{align}
&\textsf{(P0)}\; \underset{\alpha\;, [\bF_{\rf,l}]_{ij},\; M_l,\; D_l}{\textsf{maximize}}\; \sum_{k=1}^K \E\log_2(1 + \SINR_k) \\
&\quad\text{s.t.}\; \tr\( \bP (\bF_{\rf,l} \bF_{\bb,l} )^\H \bF_{\rf,l} \bF_{\bb,l}  \) + \tr(\bQ_l) \leq P_\tot \label{Eq:P1Const1}\\
&\quad\quad\;\; 2 D_l M_l \leq \cf \quad \text{for integers $D_l$ and $M_l$} \label{Eq:P1Const2}\\
&\quad\quad\;\; [\bF_{\rf,l}]_{i,j} = 1/\sqrt{N} \quad \label{Eq:P1Const3}
\end{align}\normalsize
where $1\leq l\leq L,\; 1\leq i\leq N,\; \text{and } 1\leq j \leq M_l \nn$

It is noted that the objective function is based on the received signals from the aggregate RRHs, while the constraints \eqref{Eq:P1Const1} and \eqref{Eq:P1Const2} come from per-RRH requirements. Each of these constraints comprises $L$ number of inequalities having the following meanings. The first constraint \eqref{Eq:P1Const1} implies that the per-RRH average transmit power should not exceed its maximum budget $P_\tot$, i.e. $\E||\tilde\bx_l||^2 \leq P_\tot$. The second constraint \eqref{Eq:P1Const2} indicates that the rate at which the digitally precoded symbols $\bx_l$ are forwarded from the BBU to RRH~$l$ should be no greater than the fronthaul capacity $\cf$. The last constraint \eqref{Eq:P1Const3} is a unit modulus constraint on the analog beamformers, reflecting the fact that the analog beamformers of each RRH are implemented using only phase shifters.

\subsection{Simplified Sum-Rate Maximization without Constraints}
To improve tractability, we seek a way such that \textsf{P0} is expressed as a simple objective maximization problem by merging and simplifying the constraints \eqref{Eq:P1Const1}, \eqref{Eq:P1Const2}, and \eqref{Eq:P1Const3} as follows.

\subsubsection{Per-RRH Constraint \eqref{Eq:P1Const1}}
We include this constraint into the objective function by considering its optimality
condition. More specifically, we seek to set the scaling parameter $\alpha$ to the
largest value that satisfies the power constraint for all $L$ RRHs. Clearly, the
optimum value is that which satisfies constraint \eqref{Eq:P1Const1} with equality
for the RRH $\hat{l}$ using the largest power among all RRHs. With this end, we
define two shaping matrices, a $\bar{N}\times N$ matrix $\bE_{\bar{N},l}$ and a
$\bar{M}\times N$ matrix $\bE_{\bar{M},l}$. They allow the conversion from
$\bF_{\rf}$ and $\bF_{\bb}$ to $\bF_{\rf,l}$ and $\bF_{\bb,l}$ (and vice versa),
given as follows:
\begin{align}
\bE_{\bar{N},l} &=[\b0_{N\times N(l-1)}^\T, \bI_{N}, \b0_{N\times L(l-1)}^\T]^\T\\
 \bE_{\bar{M},l} &=[\b0_{M_l\times \sum_{i=1}^{l-1}M_i }^\T, \bI_{M_l}, \b0_{M_l\times \sum_{i=l+1}^L}^\T ]^\T.
\end{align}

Applying them, we can represent \eqref{Eq:FbblDef} as
\begin{align}
\bF_{\bb,l} &=\alpha\bE_{\bar{M},l}^\T \bC^{-1} (\bH \bF_\rf)^\H  \quad\text{where} \label{Eq:Fbbl}\\
\f_{\bb,k,l}&=\alpha\bE_{\bar{M},l}^\H \bC^{-1} \bF_{\rf}^\H\bh_k. \label{Eq:fbbkl}
\end{align}
The per-RRH power constraint \eqref{Eq:P1Const1} is thereby rephrased as
\begin{align}\label{Eq:hatQ}
 \alpha^2\[\Psi_{1,l} +  \tr(\hat{\bQ}_l)\] \leq P_\tot
\end{align}
where
\begin{align}
\Psi_{1,l} &= \tr\(\bP \bH \bF_\rf \bC^{-1} \bB_{\rf,l} \bC^{-1} \bF_{\rf}^\H \bH^\H \) \label{Eq:Psi1Def}\\
\bB_{\rf,l} &= \bE_{\bar{M},l} \bF_{\rf,l}^\H \bF_{\rf,l} \bE_{\bar{M},l}^\H\\
\hat{\bQ_l} &= \bQ_l/\alpha^2 = \text{diag}\(w_{l,1}^2,\cdots, w_{l,M_l}^2 \).
\end{align}
By the definition of \eqref{Eq:tau} and \eqref{Eq:fbbkl}, $w_{m,l}^2$ is given as
\begin{align}
w_{m,l}^2 &= 3\cdot 2^{-2D_l} \sum_{k=1}^K p_k \bh_k^\H \bF_\rf \bC^{-1} \bB_{m,l} \bC^{-1} \bF_{\rf}^\H \bh_k \\
 \bB_{m,l} &= \bE_{\bar{M},l} \bE_{m} \bE_{\bar{M},l}^\H
\end{align}
where $\bE_m$ is a shaping matrix that converts $\f_{\bb,k,l}$ into $\f_{\bb,k,l,m}$, given as an $M_l\times M_l$ matrix having zeros in all entries only except for the $m$-th diagonal entry which is set as unity, e.g. $\bE_{2} = \[\begin{smallmatrix} 0& 0 \\0 & 1\end{smallmatrix}\]$ for $M_l=2$. It is also noted that the block-wise aggregation $\bF_\rf$ of $\bF_{\rf,l}$'s within $\Psi_{1,l}$ in \eqref{Eq:Psi1Def}, defined in Section~\ref{Sect:Hybrid}, can be represented as
\begin{align}
\bF_\rf = \sum_{l=1}^L \bE_{\bar{N},l} \bF_{\rf,l}\bE_{\bar{M},l}^\T.
\end{align}

The sum-rate is maximized when the RRH $\hat{l}$ generating the largest transmit power and quantization noise variance, LHS of \eqref{Eq:hatQ}, utilizes the entire power budget $P_\tot$, RHS of \eqref{Eq:hatQ}, i.e. $\Psi_{1,\hat{l}} +  \tr(\hat{\bQ}_{\hat{l}}) = P_\tot/\alpha^2 $
where
\begin{align}
\hat{l}=\arg\max_l \Psi_{1,l} + \tr(\hat{\bQ}_l). \label{Eq:hatl}
\end{align}
We accordingly apply the following replacement of \eqref{Eq:hatQ} to the objective
function of \textsf{P0}:
\begin{align}\label{Eq:alpha}
\alpha^2 = \frac{P_\tot}{\Psi_{1,\hat{l}} + \tr(\hat{\bQ}_{\hat{l}})}.
\end{align}
With this setting, $\alpha$ becomes the optimal normalization of $\bF_{\bb}$ for a given
$\bF_\rf$. Therefore, we remove the optimization parameter $\alpha$ in the
problem formulation hereafter.

\subsubsection{Fronthaul Capacity Constraint \eqref{Eq:P1Const2}} Similarly, we merge this constraint into the objective function of \textsf{P0}. We consider a fronthaul capacity limited regime where the sum-rate is maximized when utilizing the entire fronthaul capacity. This corresponds to the case when the LHS becomes as close as possible to the RHS in \eqref{Eq:P1Const2}. We thus apply the optimal number of quantization bits $D_l^* = \lfloor \cf/(2 M_l) \rfloor$ to both $\bQ_l$ in \eqref{Eq:P1Const1} and the objective function, when $M_l$ is a given value to be optimized later. Consequently, the optimization parameter $D_l$ is also removed in the problem formulation hereafter.

\subsubsection{Unit Modulus Constraint \eqref{Eq:P1Const3}} \label{Sect:UnitModulusConst}
We tentatively neglect the constraint \eqref{Eq:P1Const3} during optimization, and derive an unconstrained optimal $\bF_{\rf,l}^*$ to be specified in Section~IV. Afterwards, from the unconstrained optimal $\bF_{\rf,l}^*$, we extract the phases $\angle\bF_{\rf,l}^*$ to construct its constrained version $\bar{\bF}_{\rf,l}^* = \exp(j\angle\bF_{\rf,l})/\sqrt{N}$ abiding by \eqref{Eq:P1Const3}. The sum-rate loss brought by the different $\bF_{\rf,l}^*$ and $\bar{\bF}_{\rf,l}^*$ is clarified by simulation in Section~VI.

By utilizing the aforementioned three steps and applying the definition of $\f_{\bb,k}$ in \eqref{Eq:fbbk} to \eqref{Eq:SINR_0}, \textsf{P0} is consequently reformulated as follows.

\vspace{-10pt}\small\begin{align}
\textsf{(P1)}\; \underset{ [\bF_{\rf,l}]_{ij},\;M_l }{\textsf{maximize}}\; &\sum_{k=1}^K \E\log_2\(1 + \frac{p_k |\Psi_4|^2}{\Psi_3 + \Psi_2 + \rho^{-1}\[\tr(\hat{\bQ}_{\hat{l}}) + \Psi_{1,\hat{l}}\]}\)
\end{align}
where
\normalsize\begin{align}
\Psi_{1,l} &= \tr\(\bP \bH \bF_\rf \bC^{-1} \bB_{\rf,l} \bC^{-1} \bF_{\rf}^\H \bH^\H \), \label{Eq:Psi1}\\
\Psi_{2,k} &= \bh_k^\H \bF_\rf \hat{\bQ} \bF_\rf^\H \bh_k, \label{Eq:Psi2}\\
\Psi_{3,k} &= \bh_k^\H \bF_\rf \bC^{-1} \bF_\rf^\H \bH_{[k]}\bP_{(k)}\bH_{[k]}^\H \bF_\rf \bC^{-1}\bF_\rf^\H \bh_k, \label{Eq:Psi3}\\
\Psi_{4,k} &= \bh_k^\H \bF_\rf\bC^{-1}\bF_\rf^\H \bh_k. \label{Eq:Psi4}
\end{align}\normalsize
The matrix $\bX_{[k]}$ denotes a matrix without the $k$-th row of its original matrix $\bX$. The matrix $\bX_{(k)}$ indicates a matrix without both the $k$-th row and column of its original matrix $\bX$.

In the following subsection, we derive the large-scale approximated values of $\Psi_{1,l}$, $\Psi_{2,k}$, $\Psi_{3,k}$, $\Psi_{4,k}$, and $\tr(\hat{\bQ}_l)$ by assuming $N\rightarrow \infty$, which further simplifies \textsf{P1}.

\subsection{Large-Scale Approximated Sum-Rate Maximization}
Although the original sum-rate maximization problem \textsf{P0} is simplified as \textsf{P1} having no constraints, it is still technically challenging to solve this problem. One major difficulty is brought by the channel randomness in $\bH$ that comes from the digital beamformer's exploiting the effective channel $\bH \bF_\rf$. Another technical difficulty is incurred by the per-RRH transmit power constraint \eqref{Eq:P1Const1}, leading to the search for the RRH consuming the largest transmit power, i.e. $\hat{l}$ in \eqref{Eq:hatl}.

To circumvent these difficulties, the goal of this subsection is to approximate the objective function of \textsf{P1} by using random matrix theory \cite{Wagner:12,HoydisDebbah:13}. When the number of aggregate antennas $\bar{N}$ is sufficiently large, the following asymptotic $\SINR$ expression approximates well the exact value.

\begin{proposition} (Large-Scale Approximated $\SINR$) \emph{ As $\bar{N}\rightarrow~\infty$ with finite $\beta>0$, the received $\SINR$ at the $k$-th UE almost surely converges to a deterministic value, i.e. $\SINR_k \overset{\bar{N}\rightarrow \infty}{\rightarrow} \overline{\SINR}_k$ that is given as:
\begin{align}
\overline{\SINR}_k = \frac{ p_k {\bar{\Psi}_{4,k}}^2 }
{ \bar{\Psi}_{3,k} + \bar{\Psi}_{2,k}
 + \frac{1}{\rho} \[  \tr\(\bar{\bQ}_{\bar{l}}\) + \bar{\Psi}_{1,\bar{l}} \]  }.
\end{align}
The terms $\bar{\Psi}_{1,l}$, $\bar{\Psi}_{2,k}$, $\bar{\Psi}_{3,k}$,
$\bar{\Psi}_{4,k}$, and $\bar{\bQ}_l$ are deterministic equivalents of $\Psi_{1,l}$,
$\Psi_{2,k}$, $\Psi_{3,k}$, $\Psi_{4,k}$, and $\bQ_l$ in \textsf{P1}, given as:
\small\begin{align}
\bar{\Psi}_{1,l} &= \sum\limits_{i=1}^K \frac{p_i \tr\( \hat{\bR}_i \bT_{\bB_{\rf,l}}' \)}{\[ 1 +  \tr\(\hat{\bR}_i \bT\) \]^2 }\\
\bar{\Psi}_{2,k} &= \sum\limits_{l=1}^L  \tr\(\hat{\bR}_k \bE_{M_l,l} \bar{\bQ}_l \bE_{M_l,l}^\T \)\\
\bar{\Psi}_{3,k} &= \frac{1}{\[1 +  \tr\(\hat{\bR}_k \bT\) \]^2}\sum\limits_{i, i\neq k}^K p_i \frac{ \tr\( \hat{\bR}_i \bT_{\hat{\bR}_k }'  \)}{\[1 + \tr\(\hat{\bR}_i \bT \) \]^2} \\
\bar{\Psi}_{4,k} &= \frac{ \tr\(\hat{\bR}_k \bT \)}{1 + \tr\(\hat{\bR}_k \bT \)}\\
\bar{\bQ}_{l}&= 3\cdot 2^{-\l\lfloor\frac{\cf}{M_l}\r\rfloor}\cdot \text{diag}\(w_{l,1}^2 ,\cdots w_{l,M_l}^2 \), 
\end{align}\normalsize
where the corresponding terms are defined as follows:
\small\begin{align}
w_{l,m}^2 &= \sum_{i=1}^K \frac{p_i  \tr\(\hat{\bR}_i \bT_{\bB_{m,l}}'  \)}{\[ 1 +  \tr\(\hat{\bR}_i \bT \) \]^2  }\\
\bT&= \(\frac{1}{\bar{N}}\sum_{i=1}^K \frac{ \hat{\bR}_i}{1 + e_i} + \beta \bI_{\bar{M} }   \)^{-1} \label{Eq:T}\\
\bT_{\bB}' &= \bT\(\bB + \frac{1}{\bar{N} }\sum_{i=1}^K \frac{\hat{\bR}_i e_{i,\bB}' }{(1 + e_i)^2}  \)\bT\\
e_k &= \frac{1}{\bar{N} }\tr\(\hat{\bR}_k \bT \) \label{Eq:ek}\\
e_{k,\bB}' &= [\bee_\bB']_k\\
\bee_\bB' &= \(\bI_K - \bJ \)^{-1}\bv_{\bB}\\
[\bJ]_{ij} &= \frac{\frac{1}{\bar{N}}\tr\( \hat{\bR}_i \bT \hat{\bR}_j \bT \) }{\bar{N}(1 + e_j)^2 } \\
[\bv_{\bB}]_k &= \frac{1}{\bar{N}}\tr\( \hat{\bR_k} \bT \bB \bT  \)\\
\bB_{\rf,l} &= \bE_{\bar{M},l} \bF_{\rf,l}^\H \bF_{\rf,l} \bE_{\bar{M},l}^\H\\
\bB_{m,l} &= \bE_{\bar{M},l} \bE_{m} \bE_{\bar{M},l}^\H\\
\hat{\bR}_k &= \bF_\rf^\H\bR_k \bF_\rf\\
\bar{l}&=\arg\max_l \bar{\Psi}_{1,l} + \tr(\bar{\bQ}_l). \label{Eq:hatl_dt}
\end{align}\normalsize
\begin{proof} See Appendix.
\end{proof}
}\end{proposition}
It is noted that $e_k$ in \eqref{Eq:ek} includes its recursive expression within $\bT$ in \eqref{Eq:T}. Nonetheless, a fixed-point iteration, proved by \cite{Wagner:12,HoydisDebbah:13} and also applied in \cite{KerretGesbert:15,TabikhSlock:16,TabikhSlock:17}, guarantees fast convergence and provides $ e_k=\lim_{j\rightarrow \infty} e_k^{(j+1)}$ such that
\begin{align}
e_k^{(j+1)} = \frac{1}{\bar{N}}\tr\( \hat{\bR}_k \( \frac{1}{\bar{N}}\sum_{i=1}^K \frac{\hat{\bR}_{i}}{1 + e_i^{(j)}} + \beta \bI_{\hat{M}} \)^{-1} \) \label{Eq:eFP}
\end{align}
for $j>0$ and $e_k^{(0)}=1/\rho$. Another notable aspect is $\bar{l}$ based on the deterministic equivalents may be different from $\hat{l}$ in \eqref{Eq:hatl} due to the randomness of instantaneous CSI.

Based on the large-scale approximated $\SINR$ in \textbf{Proposition~1}, we henceforth consider the following approximated problem of \textsf{P1}:
\begin{align}
\textsf{(P2)}\; \underset{ [\bF_{\rf,l}]_{ij},\;M_l }{\textsf{maximize}}\; & \sum_{k=1}^K \log_2\(1 + \overline{\SINR}_k \)
\end{align}
where $\overline{\SINR}_k$ is given in \textbf{Proposition~1}. Compared to \textsf{P1}, the
above objective function no longer includes the expectation over channel
realizations incurred by the digital beamformer's utilizing the instantaneous
effective CSI, i.e. $\bH \bF_\rf$ in \eqref{Eq:FbbDef}, thereby simplifying the
optimization in the following sections.

\section{Analog Beamformer Design based on a Weighted Sum of Spatial Covariance Matrices}
We aim at providing an analog beamformer design criterion that relies solely on spatial covariance matrices. To provide a tractable method, we seek a way to construct $[\bF_{\rf,l}]_{ij}$ for a given $M_l$ in this section, and then optimize $M_l$ under the given $[\bF_{\rf,l}]_{ij}$ in Section~V.

Within this section, for simplicity we consider the following assumptions. First, we neglect quantization noise impact on $[\bF_{\rf,l}]_{ij}$ decision by considering $\cf\rightarrow~\infty$. The limited fronthaul constraint is, instead, re-incorporated in the selection of the optimal values for $M_l$ in Section~V. Our approach is justified by the fact that $M_l$ is the number of data streams to be forwarded from the BBU to RRH $l$, which has a much more dominant effect in the fronthaul capacity constraint compared to the specific analog beamformer $\bF_{\rf,l}$ used. Second, we neglect initially the unit modulus constraints \eqref{Eq:P1Const3} on the entries of the analog beamformers, which are enforced only after the optimal $M_l$ and unconstrained $\bF_{\rf,l}$ have been found. Third, we consider a single RRH, i.e. $L=1$, with the intention of enabling independent analog beamformer design for different RRHs. This is partly justified by the fact that the channels for different RRHs are uncorrelated. Last, we consider ZF digital beamformer and unit transmit power allocation per UE, i.e. $\beta=0$ in \eqref{Eq:FbbDef} and $p_k=1$, which provides more design intuition by analysis. All these simplifications are to be compensated by optimizing the RF chain activations and digital beamformer design in Section~V.

\subsection{Motivation -- Equal Combining Approach \emph{\cite{ParkHeath:17}} }
The method used in our analog beamformer design stems from \cite{ParkHeath:17}. The objective in this prior work is to maximize average signal-to-leakage-plus-noise ratio ($\mathsf{SLNR}$) under an RZF digital beamformer. For this purpose, it first generates \emph{equal combining} of all the spatial covariance matrices of UEs, and then constructs the analog beamformer containing the columns selected by the eigenvectors associated with the largest eigenvalues up to $M_l$ number of RF chains.

The effectiveness of such a $[\bF_\rf]_{ij}$ construction based on equally combined spatial covariance matrices hinges on the UE channel conditions, namely identically distributed $\bR_k$'s over UEs. This channel condition holds probably when all UEs are concentrated in a hotspot while being served by a single RRH or multiple colocated RRHs. It leads to identical large-scale fading, i.e. the same path losses $\delta_k$'s, and i.i.d. small-scale fading, resulting in the desired i.i.d. $\bR_k$'s. On the other hand, when RRHs and/or UEs are more distributed, the channel path losses can hardly be identical. This breaks the i.i.d. $\bR_k$ channel condition, thus urging another method of combining spatial covariance matrices.

\begin{figure}[t]
	\centering
		\includegraphics[width=\linewidth]{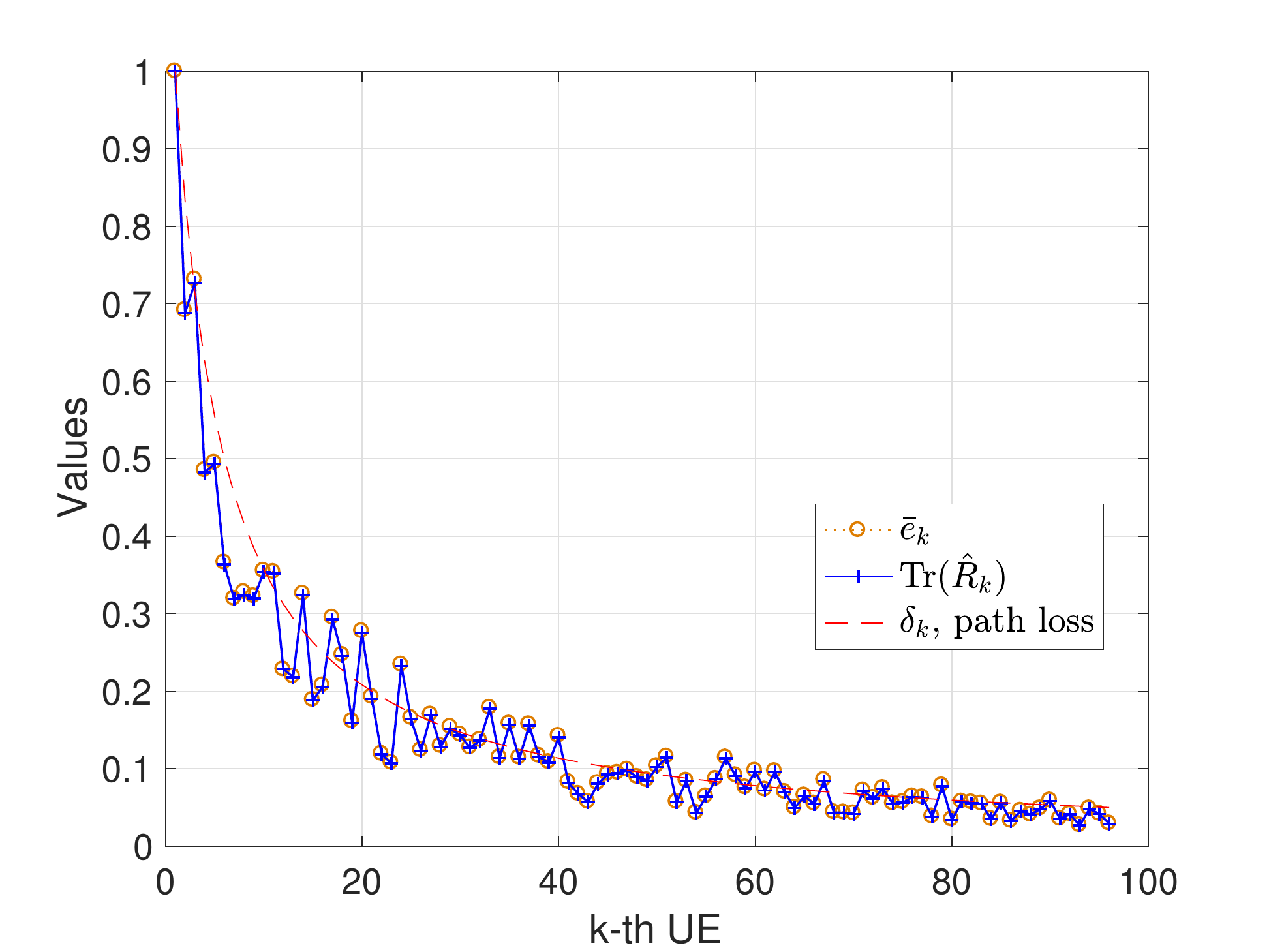}
	\caption{Normalized $\bar{e}_k/\max_k \{\bar{e}_k\}$ compared to $\tr(\hat{\bR}_k)$ with respect to UE $k$ associated with its normalized path loss $\delta_k/\max_k\{\delta_k\}$ ($N=128$, $K=100$, $L=1$, $\delta_k\in[0.01,1]$ equally-spaced in descending order).}
	\label{F:sumrate_rzf_vs_rfc}
\end{figure}

From a theoretical point of view, the said discussion can be clarified by the large-scale approximated average (or sum) $\textsf{SLNR}$ in \cite[Eq. (11)]{ParkHeath:17} iterated as follows.
\begin{align}
\mathsf{SLNR} \overset{N\rightarrow \infty}{\rightarrow}& \sum_{k=1}^K \tr\(\hat{\bR}_k \bar{\bT} \) \label{Eq:SLNR}\\
\text{where }\bar{\bT}=& \[\frac{1}{N}\sum_{i=1}^K \frac{\hat{\bR}_i}{\bar{e}_i} +  \bI\]^{-1} \label{Eq:barT}\\
\overset{\text{if } \bar{e}_i = \bar{e}_1}{=}& \[ \frac{1}{N \bar{e}_1}\sum_{i=1}^K \hat{\bR}_i +  \bI\]^{-1} \label{Eq:eAss}
\end{align}
So long as \eqref{Eq:eAss} holds under $\bar{e}_i=\bar{e}_1$ $\forall i=1,\cdots, K$, it has been shown by \cite[Proposition~2]{ParkHeath:17} that the $[\bF_\rf]_{ij}$ construction based on equal combining of spatial covariance maximizes the large-scale approximated $\mathsf{SLNR}$, i.e. Eq. \eqref{Eq:SLNR}.

Unfortunately, \eqref{Eq:eAss} becomes no longer valid under different path losses $\delta_i$'s since $\bar{e}_i$'s cannot be identical as illustrated in Fig.~2, showing the $\bar{e}_i$'s (dotted orange circle) and $\delta_i$'s (dashed red) normalized respectively by their maximum values are strongly correlated. This motivates us to design an analog beamformer based on a weighted sum of spatial covariance matrices, described in the next subsection.

Note that $\bar{\bT}$ in \eqref{Eq:barT} under ZF is rephrased from the original expression with RZF after minor modifications provided in \cite{Wagner:12,WagnerSPAWC:10}. A major change is the replacement of $1+\bar{e}_i$ in the denominator by $\bar{e}_i$ of which the value can still be found via a fixed-point iteration algorithm as in~\eqref{Eq:eFP}. Another notable difference is $\bI$ within $\bar{\bT}$ in \eqref{Eq:barT}, which is $\beta \bI$ in the original expression under RZF.

\subsection{Trace-Weighted Approach}

As discussed in the previous subsection, it is difficult to cope with the different path losses of UEs through the analog beamformer based on equal combining of spatial covariance matrices. For this reason, we seek an analog beamformer design based on a weighted sum of spatial covariance matrices, and focus on the weighting factor decision.

The key idea starts from maximizing the large-scale approximated $\SINR$ given by \cite{Wagner:12} as
\begin{align}
\SINR &\overset{N\rightarrow \infty}\rightarrow \(\sum_{k=1}^K \frac{1}{\tr\( \hat{\bR}_k \bar{\bT} \)} \)^{-1}\\
 &\quad\leq  \frac{1}{K^2}\sum_{k=1}^K \tr\( \hat{\bR}_k \bar{\bT} \) \label{Eq:ZFupper}
\end{align}
where the last step follows from the harmonic-arithmetic mean inequality \cite{PolyaBook:Inequalities}, i.e. $K/(\sum_{k=1}^K 1/x_k )\leq \sum_{k=1}^K x_k/K$ for $x_k~=~\tr(\hat{\bR}_k\bar{\bT})$ where the equality holds for identical $x_k$'s. We hereafter focus on the upper bound \eqref{Eq:ZFupper} as the objective function.

Applying the definition $\bR_k = \delta_k \mathbf{\Theta}_k$ in \eqref{Eq:RklDef} with an additional definition of the effective fast-fading covariance matrix $\hat{\mathbf{\Theta}}_k = \bF_\rf^\H \mathbf{\Theta}_k \bF_\rf $, the objective function to be maximized becomes
\begin{align}
\sum_{k=1}^K \tr\(\hat{\bR}_k \bar{\bT}\) = \tr\(\sum_{k=1}^K \hat{\bR}_k\[\frac{1}{N}\sum_{i=1}^K \frac{\hat{\bR}_i}{\bar{e}_i} + \bI\]^{-1}  \)&\\
\quad \quad\quad\quad= \tr\(\sum_{k=1}^K \delta_k \hat{\mathbf{\Theta}}_k \[\frac{1}{N}\sum_{i=1}^K \frac{\delta_i \hat{\mathbf{\Theta}}_i}{\bar{e}_i} + \bI \]^{-1}  \)&. \label{Eq:ZFupper2}
\end{align}

Next, we aim at making \eqref{Eq:ZFupper2} become similar to \eqref{Eq:eAss} that includes equal combining of i.i.d. spatial covariance matrices, i.e. $\sum_{i} \hat{\bR}_i$, thereby allowing the analog beamformer design based simply on the (weighted) sum of spatial covariance matrices. With this end, we consider the following assumption motivated by the simulation result shown in Fig.~2.
\begin{assumption} Let $\bar{e}_1= \max_k \{\bar{e}_k\}$ and $\delta_1 = \max_k\{\delta_k\}$. Then, $\bar{e}_k/\bar{e}_1$ is identical to $\delta_k/\delta_1$ $\forall k=1,\cdots,K$.
\end{assumption}

As long as Assumption~1 holds, we observe our objective function becomes
\begin{align}
\eqref{Eq:ZFupper2}&= \tr\(\sum_{k=1}^K \delta_k \hat{\mathbf{\Theta}}_k \[\frac{\delta_1}{N \bar{e}_1}\sum_{i=1}^K \hat{\mathbf{\Theta}}_i +\bI \]^{-1}  \)\\
&\leq \delta_1\tr\(\sum_{k=1}^K  \hat{\mathbf{\Theta}}_k \[\frac{\delta_1}{N \bar{e}_1}\sum_{i=1}^K \hat{\mathbf{\Theta}}_i +\bI \]^{-1}  \). \label{Eq:ZFThetak}
\end{align}

It is remarkable that the effective fast-fading covariance matrices $\hat{\mathbf{\Theta}}_k$'s in \eqref{Eq:ZFThetak} are i.i.d. over UEs. Therefore, equal combining of $\hat{\mathbf{\Theta}}_k$ maximizes \eqref{Eq:ZFThetak}, proven trivially by reiterating \cite[Proposition 2]{ParkHeath:17}, i.e. applying eigenvalue decomposition to $\hat{\mathbf{\Theta}}_k$ and Cauchy's interlacing theorem~\cite{HaemersCauchyInterlacing:95}.

Recalling the definition \eqref{Eq:RklDef}, we can replace $\hat{\mathbf{\Theta}}_k$ by $\hat{\bR}_k/\delta_k$ that can approximately be exchanged with $\hat{\bR}_k/\tr(\hat{\bR}_k)$, on the basis of the strong correlation between $\delta_k$ and $\tr(\hat{\bR}_k)$ as shown by the simulation in Fig.~2. This consequently provides the following analog beamformer design based on the \emph{trace-weighted} combining of spatial covariance matrices.

\begin{remark} \emph{When $L=1$ with $p_k=1\;\forall k$, as $\bar{N},\cf\rightarrow \infty$ for $\beta= 0$ with Assumption~1, the upper bound of $\overline{\SINR}_k$ is maximized by $\bF_\rf$ having its columns  composed of the $M_l$ eigenvectors associated with the $M_l$ largest eigenvalues of $\hat{\bR}_l = \sum_{k=1}^K \hat{\bR}_{k,l}/\tr\(\hat{\bR}_{k,l}\)$.}\\
\end{remark}

One major benefit of our proposed analog beamformer design is its sole dependency on spatial covariance matrices, regardless of $\delta_k$ that may need additional UE location information, as well as of $\bar{e}_k$ requiring an iterative search used in \eqref{Eq:eFP}. As in the equal combining method \cite{ParkHeath:17}, the proposed design is thus free from instantaneous CSI, while additionally coping with different path loss channels by still using spatial covariance matrices.

It is remarkable that for the same path losses of UEs, our trace-weighted approach for analog beamformer design straightforwardly becomes the approach based on equal combining of spatial covariance matrices, although they are supposed to maximize different objective functions $\SINR$ and $\mathsf{SLNR}$ under different ZF and RZF digital beamformer setups respectively. It is also interesting to mention that proving the trace-weighted analog beamforming criterion can also be achieved by a similar approach provided in \cite{HeathJSTSP:16}; namely, minimizing the Frobenious distance between the ZF hybrid precoding matrix and its ideal ZF precoding matrix relying respectively on the effective CSI $\bH \bF_\rf$ and the exact CSI $\bH$, omitted and replaced by the aforementioned derivation.

Lastly, it is noted that $\bar{e}_k$, required to prove the $\bF_\rf$ decision, is not only relying on an iterative search but also is a function of $\bF_\rf$ itself. Due to this recursive nature, it is difficult to prove Assumption~1 by analysis. To detour the problem, we resort to simulation illustrated in Fig.~2, where the proposed trace-weighted $\bF_\rf$ construction is used.

In the following section, the trace-weighted analog beamformer design is to be combined with optimizing the RF chain activation that affects the size of $\bF_\rf$. Their aggregate effectiveness is to be validated by simulation under a more general environment in Section~VI.

\section{Hybrid Precoding Design under Limited Fronthaul Capacity}
The objective of this section is to propose a hybrid precoding design for \textsf{P0}, which namely optimizes: (i) fronthaul compression level $D_l$, (ii) RF chain activation $M_l$, (iii) analog beamformer construction $[\bF_{\rf,l}]_{ij}$, and (iv) digital beamformer amplitude scaling factor $\alpha$ for a given transmit power constraint per RRH. The design is based on spatial covariance matrices (\textbf{Remark~1} in Section~IV) and large-scale approximated $\SINR$ (\textbf{Proposition~1} in Section~III), summarized in \textbf{Algorithm~1} and described in the following subsections.

\subsection{Analog Beamformer Design}
In C-RAN architecture, all the analog beamformers of RRHs are determined at the BBU. Such decisions rely on jointly optimizing the size of $\bF_{\rf,l}$ and the entities of $\bF_{\rf,l}$, i.e. $M_l$ and $[\bF_{\rf,l}]_{ij}$, revealing the following technical challenges.

One major difficulty comes from the optimal decision of $M_l$, which affects not only the size of the analog beamformer $\bF_{\rf,l}\in\mC^{N\times M_l}$ but also the size of the digital beamformer $\bF_{\bb}\in^{\bar{M}\times K}$ for $\bar{M}=\sum_{l=1}^L M_l$. Therefore, adjusting $M_l$ affects the quantization noise variance $\bQ_l$ since fronthaul compression is performed after applying $\bF_{\bb}$ as described in Section~\ref{Sect:SysTransmit}. Consequently, it is difficult to anticipate the impact of adjusting $M_l$ on our objective function, sum-rate. Another difficulty originates from the per-RRH transmit power constraints in \eqref{Eq:P1Const1}. Provided that $\bF_\bb$ is normalized by a single quantity $\alpha$ at the BBU, it requires a search for the $\hat{l}$-th RRH producing the largest transmit power for a given channel and quantization noise.

For these reasons, it is difficult to solve the original joint optimization problem of $M_l$ and $[\bF_{\rf,l}]_{ij}$. To detour this, we propose a sub-optimal algorithm that sequentially optimizes $[\bF_{\rf,l}]_{ij}$ and $M_l$, while exploiting spatial covariance matrices (\textbf{Remark~1}) and the large-scale approximated $\SINR$ (\textbf{Proposition~1}), described next.

\subsubsection{Construction of $[\bF_{\rf,l}^*]_{ij}$}
Suppose a given $M_l\leq \hat{M}$. For this given value, $[\bF_{\rf,l}]_{ij}$ is constructed based on the trace-weighted approach in \textbf{Remark~1}.

\subsubsection{Selection of $M_l^*$}
After the construction of $[\bF_{\rf,l}]_{ij}$ for each $M_l\in\{1,2,\dots,\hat{M}\}$, we find the optimal $M_l^*$ that maximizes the sum-rate of \textsf{P2} by treating $\overline{\SINR}_k$ in \textbf{Proposition~1} as the approximation of its instantaneous $\SINR_k$.

\subsubsection{Unit modulus constraint on $\bF_{\rf,l}$}
Up to this point, the unit modulus constraint in \eqref{Eq:P1Const3} has been neglected. This constraint is re-introduced after the selection of $M_l^*$ by simply setting the entries of the analog beamformers to have unit modulus and the same phases as the unconstrained beamformer, i.e. we select $\bar{\bF}_{\rf,l}^* = \exp(j\angle\bF_{\rf,l}^*)/\sqrt{N}$ as described in Section~\ref{Sect:UnitModulusConst}.

The aforementioned procedures are summarized as Step~1 in \textbf{Algorithm~1}.

\begin{algorithm}[bt]
    \caption{Find $M_l^*$, $\bar{\bF}_{\rf,l}^*$ and $\alpha^*$ in $\bF_{\bb}$}
    \begin{algorithmic}[1]
        \Statex Step 1 -- Find $M_l^*$ and $\bar{\bF}_{\rf,l}^*$ with input: $\bR_{k,l}$
        \For{$l=1:L$}
        \For{$M_l=1:\hat{M}$}
        \State $\bar{\bR}_{k,l}= \bR_{k,l}/\tr(\bR_{k,l})$
        \State $\bar{\bR}_{l}= \sum_{k=1}^K \bar{\bR}_{k,l}$
        \State $\bF_{\rf,l}$'s columns $\leftarrow $ $M_l$ largest eigenvectors of $\bar{\bR}_l $
        \EndFor
        \EndFor
        \State $\{M_l^* : l=1,\cdots,L\}$
        \Statex $=\underset{\{M_l:l=1,\cdots,L\}}{\arg\max}\; \sum_{k=1}^K \log_2(1+ \overline{\SINR}_k)$
        \State ${\bF}_{\rf,l}^* \leftarrow$ $M_l^*$ column vectors of ${\bF}_{\rf,l}$
        \State $\bar{\bF}_{\rf,l}^* = \exp(j\angle\bF_{\rf,l} )/\sqrt{N}$
        \Statex
        \Statex Step 2 -- Find $\alpha^*$ in $\bF_{\bb}$ with input: $\bar{\bF}_{\rf}^*$ and $\bH \bar{\bF}_{\rf}^*$
        \State $\hat{l}=\arg\max_l\; \Psi_{1,l} + \tr(\hat{\bQ}_l)$
        \State $\SINR_k \leftarrow \hat{l}, \bar{\bF}^*_\rf$
        \State $\alpha^*\leftarrow \arg\max_\alpha\; \sum_{k=1}^K \E[\log_2(1 + \SINR_k)]$
    \end{algorithmic}
\end{algorithm}

\subsection{Digital Beamformer Design}
For the given analog beamformer $\bar{\bF}_\rf^*$ decided by the deterministic quantities of spatial covariance matrices and the large-scale approximated sum-rate, we investigate the optimal RZF beamformer based on the instantaneous effective CSI $\bH \bar{\bF}_\rf^*$. The parameter to be optimized is the $\bF_\bb$'s scaling parameter $\alpha$ such that it guarantees each RRH's transmit power constraint \eqref{Eq:P1Const1}. Note that this parameter affects all the RRHs identically, although each RRH has a different channel and quantization noise.

For this reason, we consider RRH $\hat{l}$ inducing the largest transmit power and quantization noise, and set the optimal $\alpha$ as the value making RRH $\hat{l}$ use the entire transmit power budget $P_\tot$, expressed in \eqref{Eq:alpha}. The selection of $\hat{l}$ out of $L$ RRHs is thereby given as $\hat{l}=\arg\max_l \Psi_{1,l} + \tr(\hat{\bQ}_l)$, described as Step~2 in \textbf{Algorithm~1}.

\section{Numerical Evaluation}

In this section, we evaluate the performance of \textbf{Algorithm~1} by means Monte Carlo
simulations. In the simulated system, we consider each of the $L$ RRHs is equipped
with a uniform linear array with $N$ antenna elements with inter-element distance
of half wavelength, whereas each of the $K$ UEs has a single antenna. The transmit
power budget $P_\tot$ of each RRH is $30$~dBm. The average noise floor $\sigma^2$ at
each UE is set as $-116~$dBm. For the generation of channel responses, we consider a
multi-path channel model with $32$ multi-path components. To each of the multipath
components, we associate an angle of departure $\phi_i$ of the $i$-th path for $i=1,\cdots,32$, which is uniformly distributed over $[0,2\pi]$. The complex gain $\alpha_i$ of each path
follows a zero mean complex Gaussian distribution, with the aggregate variance of
all paths being equal and normalized as unity. For ease of exposition, we consider
each RRH has equal distance to UE~$k$, i.e. $d_{k,l}=d_k$. Given this, the channel
response vector from RRH $l$ to user $k$ reads $\bh_{k,l} = {d_{k,l}}^{-\eta/2}\sum_{i=1}^{32}{\alpha_{i}}{\ba_i\(\phi_i\)}$, where ${\ba_i\(\phi_i\)}$ is the array response for the $i$-th path given by ${\ba_i\(\phi_i\)} = [1, e^{-j\pi\cos(\phi_i)},\cdots,e^{-j\pi(N-1)\cos(\phi_i)} ]^\T$ and its spatial covariance matrix is $\bR_{k,l} = \E[\bh_{k,l} \bh_{k,l}^\H]$. As
can be easily verified, the above channel model  is a special case of the general
model presented in \eqref{Eq:hklDef} and \eqref{Eq:RklDef}. In the implementation of
\textbf{Algorithm~1} used for the simulations, we restrict the number of active RF chains to
be identical in all RRHs, i.e. we enforce $M_l=~M\;\; \forall l$. This choice
simplifies significantly the maximization step in line 8 of \textbf{Algorithm~1}, and will be
numerically justified later in this section.

\begin{figure}[t]
	\centering
		\includegraphics[width=\linewidth]{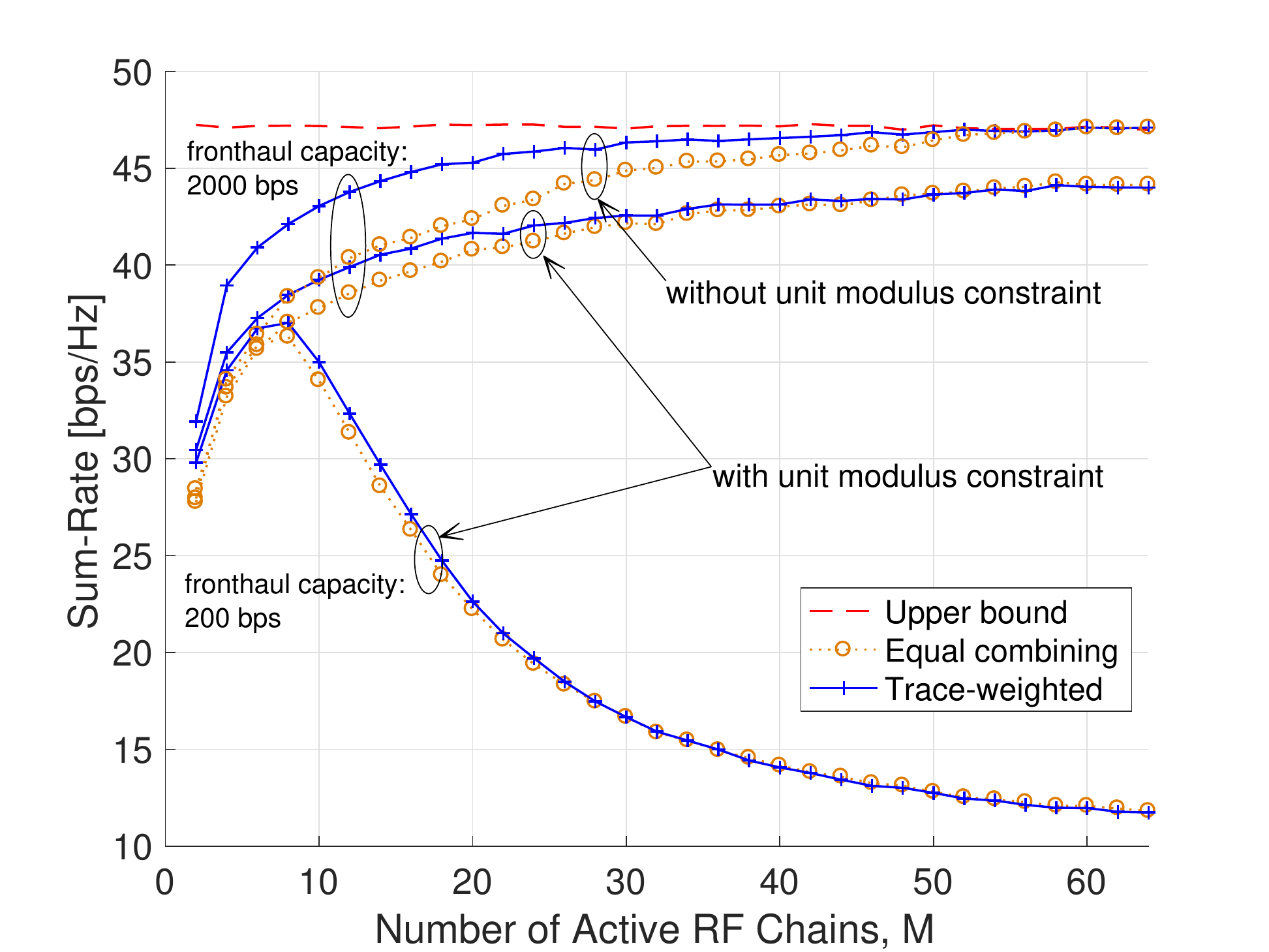}
	\caption{Sum-rate with respect to the number of active RF chains ($N=64$, $K=3$, $L=2$, $d_{1,1}=d_{1,2}=1000$~m, $d_{2,1}=d_{2,2}=500$~m, $d_{3,1}=d_{3,2}=100$~m).}
	\label{F:sumrate_rzf_vs_rfc}
\end{figure}

First, we evaluate the performance of the proposed method for design of the analog beamformer as a function of the number $M$ of active RF chains per RRH. In Fig.~\ref{F:sumrate_rzf_vs_rfc} the sum-rate obtained by our proposed method (trace-weighted, represented as {\color{blue}{solid blue~+}}) is compared to that of a similar method which applies equal weighting to the covariance of all users (equal combining\footnotemark, {\color{orange}{dotted orange~$\circ$}}) for two different fronthaul capacities $C_F = \{200, 2000\}$. \footnotetext{More specifically, line 3 of \textbf{Algorithm~1} is replaced by $\bar{\bR}_{k,l}=\bR_{k,l}$.}As an upper bound, the sum-rate performance of a fully-digital RZF precoder based on instantaneous CSI and operating over fronthauls with unlimited capacity is also presented. Our proposed design outperforms the design with equal combining in terms of sum-rate, especially when the number of active RF chains is low. In addition, the figure illustrates the impact of RF chain activation on the sum-rate. For sufficient fronthaul capacity $\cf=2000$~bps, sum-rate is a monotone increasing function of the active number of RF chains $M$, even after exceeding the number of UEs as also observed in \cite{ParkHeath:17}. In this case, activating the whole set of RF chains always provides the highest sum-rate, i.e. $M^* = \hat{M}$. For insufficient fronthaul capacity $\cf=200$~bps, on the contrary, the quantization noise variance after fronthaul compression severely grows as $M$ increases, and thus activating only a subset of the available RF chains better improves sum-rate. We also evaluate the impact of enforcing the unit modulus constraint in line 10 of \textbf{Algorithm~1} by comparing the performance of the proposed method to that of one which does not enforce the unit modulus constraint. As can be observed, the unconstrained algorithm reaches the performance of the bound for sufficiently large $M$, while the unit-modulus design has a small but significant loss in comparison. The proposed design could be further improved by an additional optimization such as a compensation matrix method in \cite{ParkHeath:17}, which is an interesting topic for future work.

\begin{figure}[t]
	\centering
		\includegraphics[width=\linewidth]{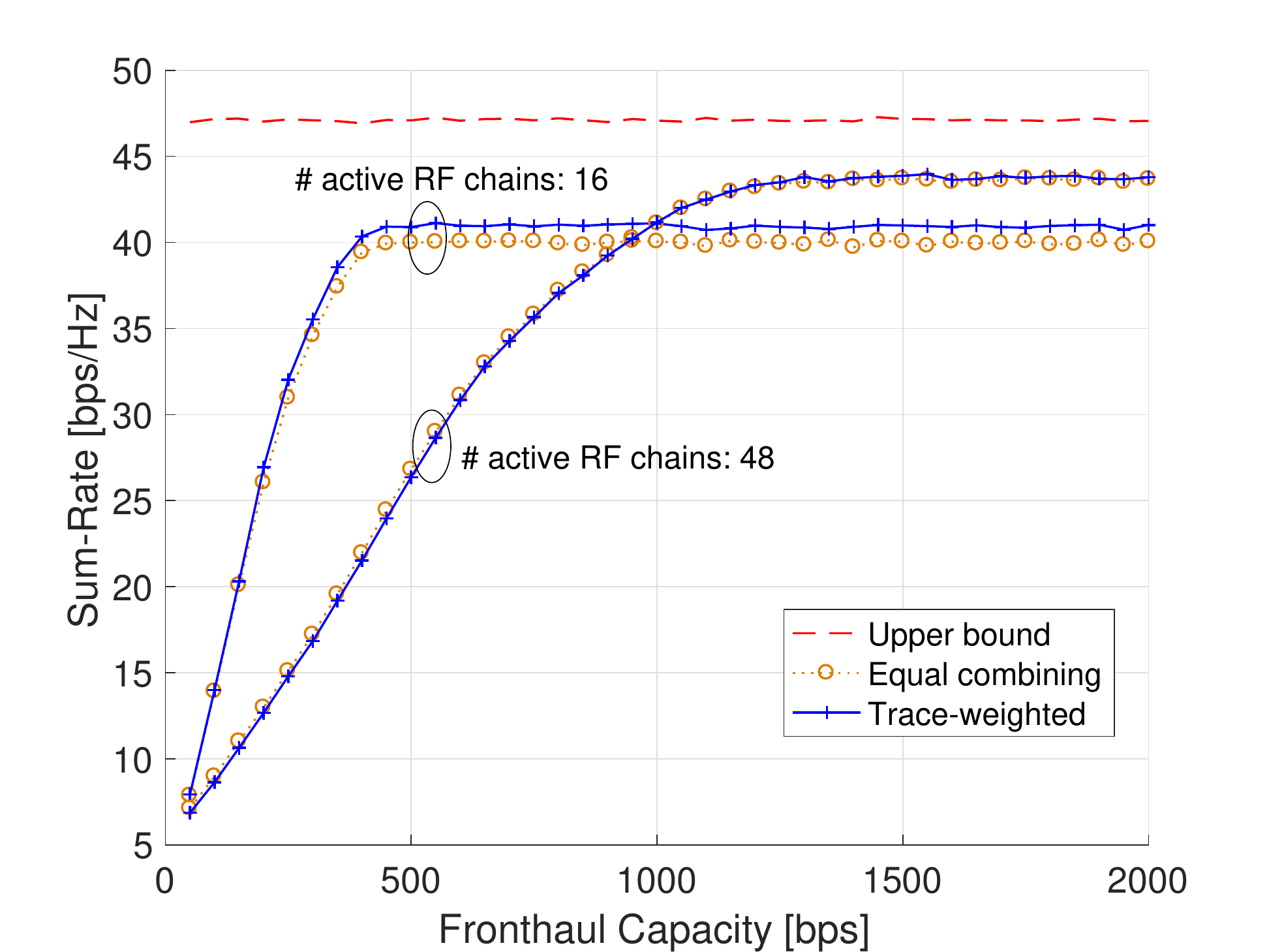}
	\caption{Sum-rate with respect to fronthaul capacity ($N=64$, $K=3$, $L=2$, $d_{1,1}=d_{1,2}=1000$~m, $d_{2,1}=d_{2,2}=500$~m, $d_{3,1}=d_{3,2}=100$~m).}
	\label{F:sumrate_rzf_vs_fc}
\end{figure}

Fig.~\ref{F:sumrate_rzf_vs_fc} evaluates the effectiveness of the proposed trace-weighted combining method with respect to the number of active RF chains and fronthaul capacity. For large number of active RF chains fixed as $M=48$, both trace-weighted and equal combining methods provide non-distinguishable sum-rates. For small number of active RF chains fixed as $M=16$, on the other hand, the trace-weighted scheme leads to higher sum-rate compared to that of the equal combining. Based on this result for $M=16$, we can also expect that the trace-weighted scheme will provide higher sum-rate gain under severe limitations on (i) fronthaul capacity and (ii) the number of total RF chains per RRH. As can be seen, a small number of active RF chains provide better sum-rate performance for fronthauls with low capacity. This is explained by the fact that, with low number of RF chains, the amount of information forwarded from the BBU to each RRH is smaller and can be quantized more effectively. With $M=48$, on the other hand, quantization noise severely degrades the sum-rate under low fronthaul capacity. It is also observed that our proposed trace-weighted design obtains larger performance advantage over equal combining when the number of active RF chains is low.

\begin{figure}[t]
\centering
\subfloat[Sum-rate with optimal RF chain activation]{
\includegraphics[width=\linewidth]{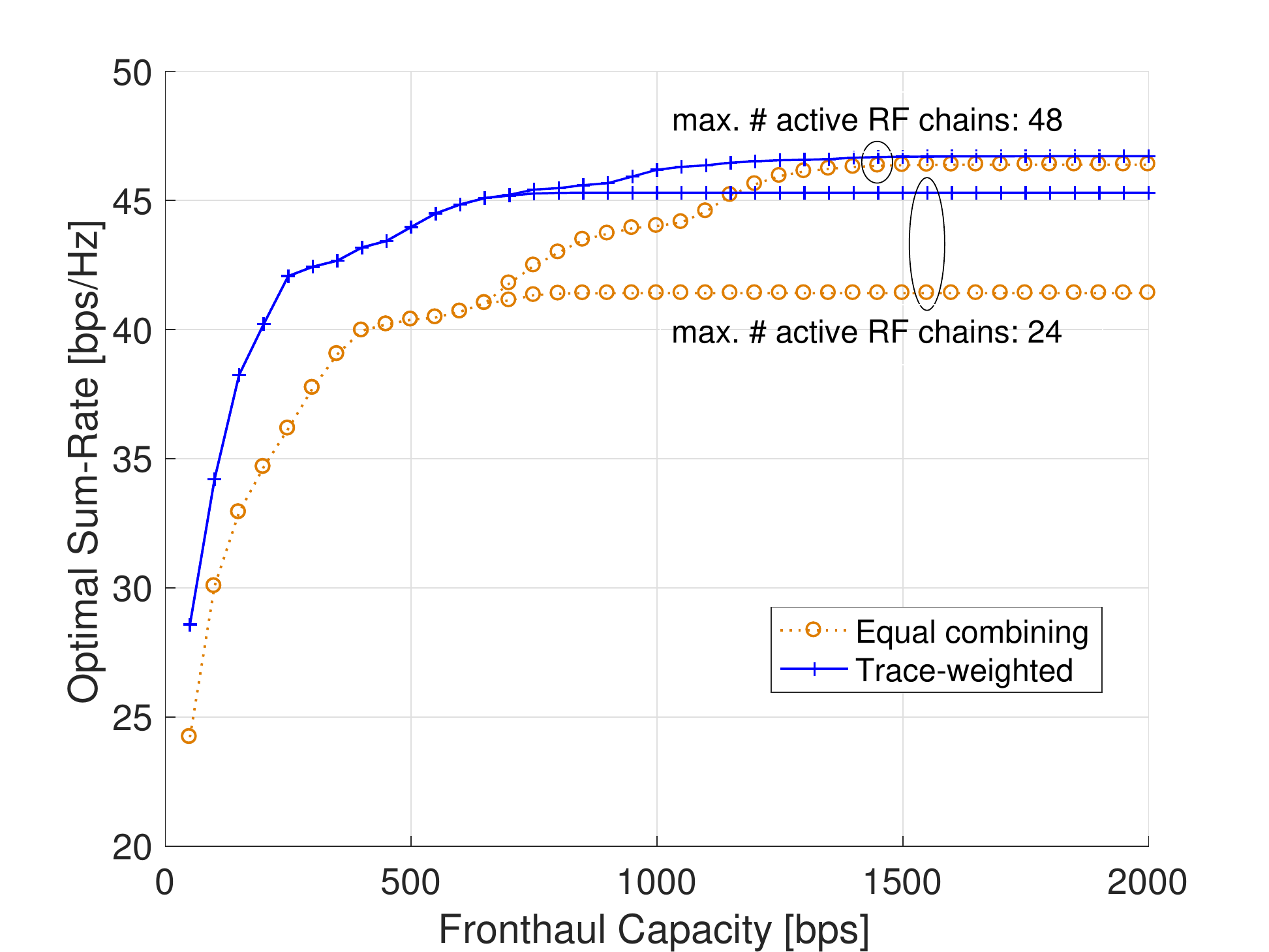}
\label{F:optsumrate_rzf_vs_fc}} \\
\subfloat[Sum-Rate with and without optimal RF chain activation]
{\includegraphics[width=\linewidth]{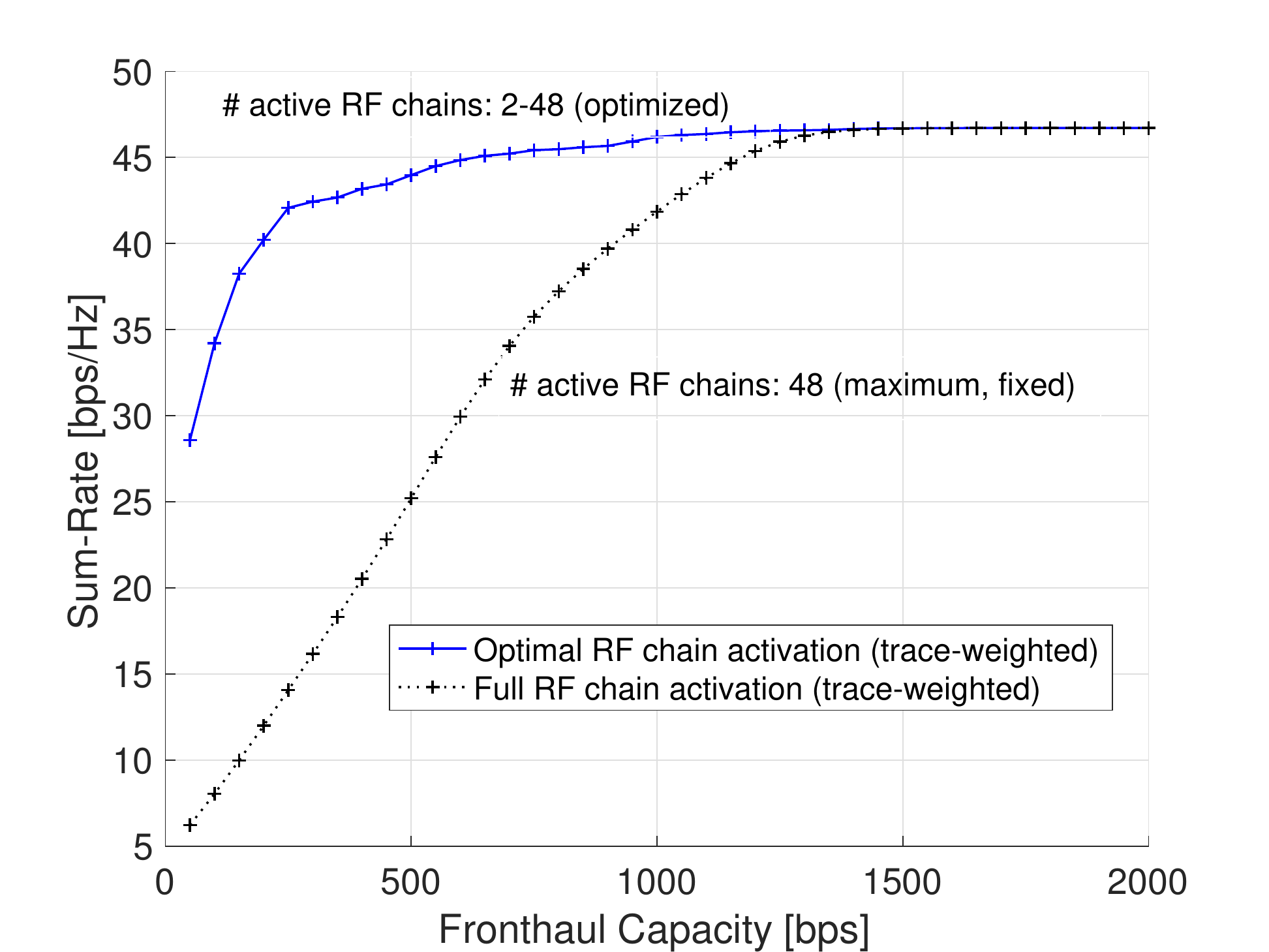}
\label{F:optimal_rzf_vs_fc}}
\caption{Sum-rate with the RF chain activation optimized by \textbf{Algorithm~1}, with respect to fronthaul capacity ($N=64$, $K=3$, $L=2$, $d_{1,1}=d_{1,2}=1000$~m, $d_{2,1}=d_{2,2}=500$~m, $d_{3,1}=d_{3,2}=100$~m).}
\label{F:optimal}
\end{figure}

\begin{figure}[t]
\centering
\subfloat[Optimal number of active RF chains.]{
\includegraphics[width=\linewidth]{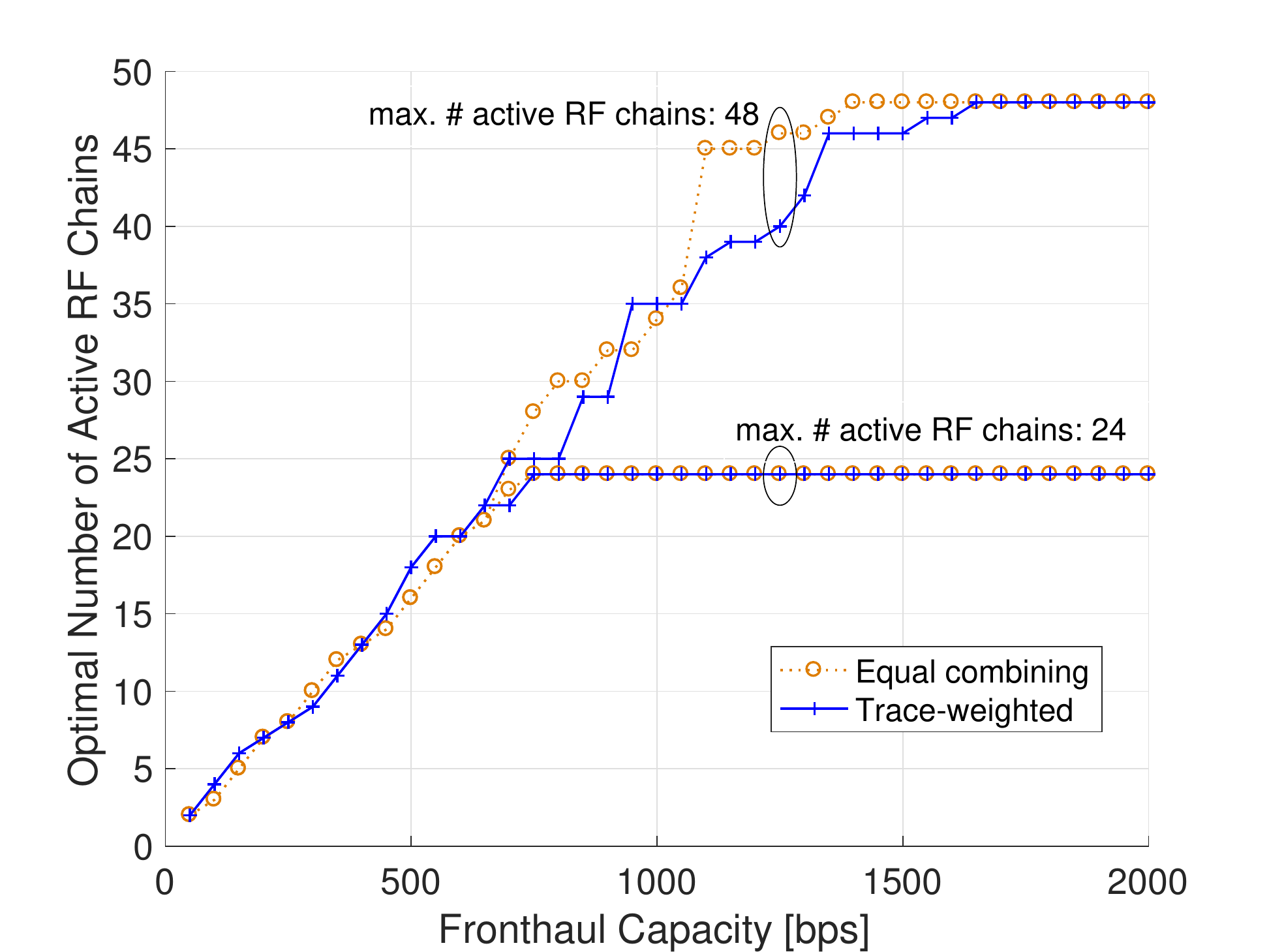}
\label{F:optrfc_rzf_vs_fc}}\\
\subfloat[Optimal sum-rate.]{
\includegraphics[width=\linewidth]{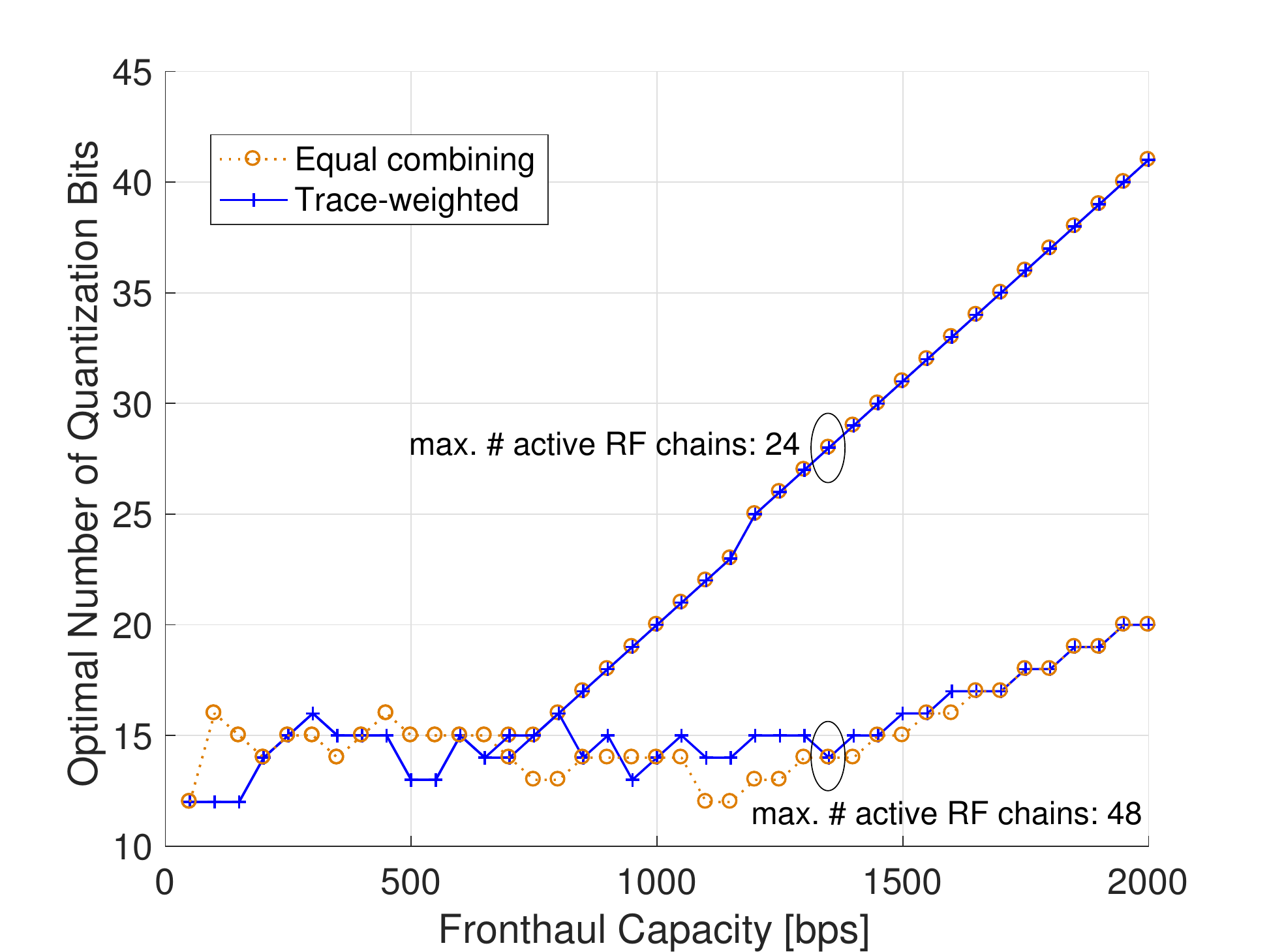}
\label{F:opt_fixed_D_vs_fc}}
\caption{Sum-rate with the number of active RF chains optimized by \textbf{Algorithm~1}, with respect to fronthaul capacity ($N=64$, $K=3$, $L=2$, $d_{1,1}=d_{1,2}=1000$~m, $d_{2,1}=d_{2,2}=500$~m, $d_{3,1}=d_{3,2}=100$~m).}
\label{F:optrfcD}
\end{figure}

In Fig.~\ref{F:optimal} we present the sum-rate performance of the full optimization process in \textbf{Algorithm~1}, where the hybrid precoder is optimized jointly with the number of active RF chains~$M$. As seen in Fig.~\ref{F:optsumrate_rzf_vs_fc}, where the total number of available RF chains per RRH has been set to $\hat{M}=\{24,48\}$, the proposed trace-weighed method provides higher sum-rate under more limited fronthaul capacity as well as under smaller number of total RF chains per RRH. This is consistent with our findings on Fig.~\ref{F:sumrate_rzf_vs_fc}. In addition, Fig.~\ref{F:optimal_rzf_vs_fc} emphasizes the importance of appropriate RF chain activation under limited fronthaul capacity. Full RF chain activation crumbles in effectiveness when fronthaul capacity is small due to severe quantization noise variance brought by coarse quantization levels. With our proposed joint optimization of the hybrid precoder design and the number of active RF chains, the effect of quantization noise is largely mitigated, and moderate-to-high sum-rates can be obtained even with severe limitation of the fronthaul capacity.

In Fig.~\ref{F:optrfcD}, we analyze the effect of the fronthaul capacity on the optimal numbers of active RF chains $M$ and quantization bits $D_l$ provided by \textbf{Algorithm~1} for two cases of maximum number of available RF chains $\hat{M}=\{24,48\}$. Fig.~\ref{F:optrfc_rzf_vs_fc} reveals that the number of active RF chains grows linearly with the fronthaul capacity as long as the maximum number of RF chains is not exceeded. On the other hand, Fig.~\ref{F:opt_fixed_D_vs_fc} shows that the number of quantization bits is kept approximately constant with increasing $C_F$ as long as there are still RF chains available. After all available RF chains are in use ($M=\hat{M}$), further fronthaul capacity is used to quantize more finely the symbols forwarded through the fronthaul.

\begin{figure}[t]
    \hspace{-.6cm}    \includegraphics[width=10cm]{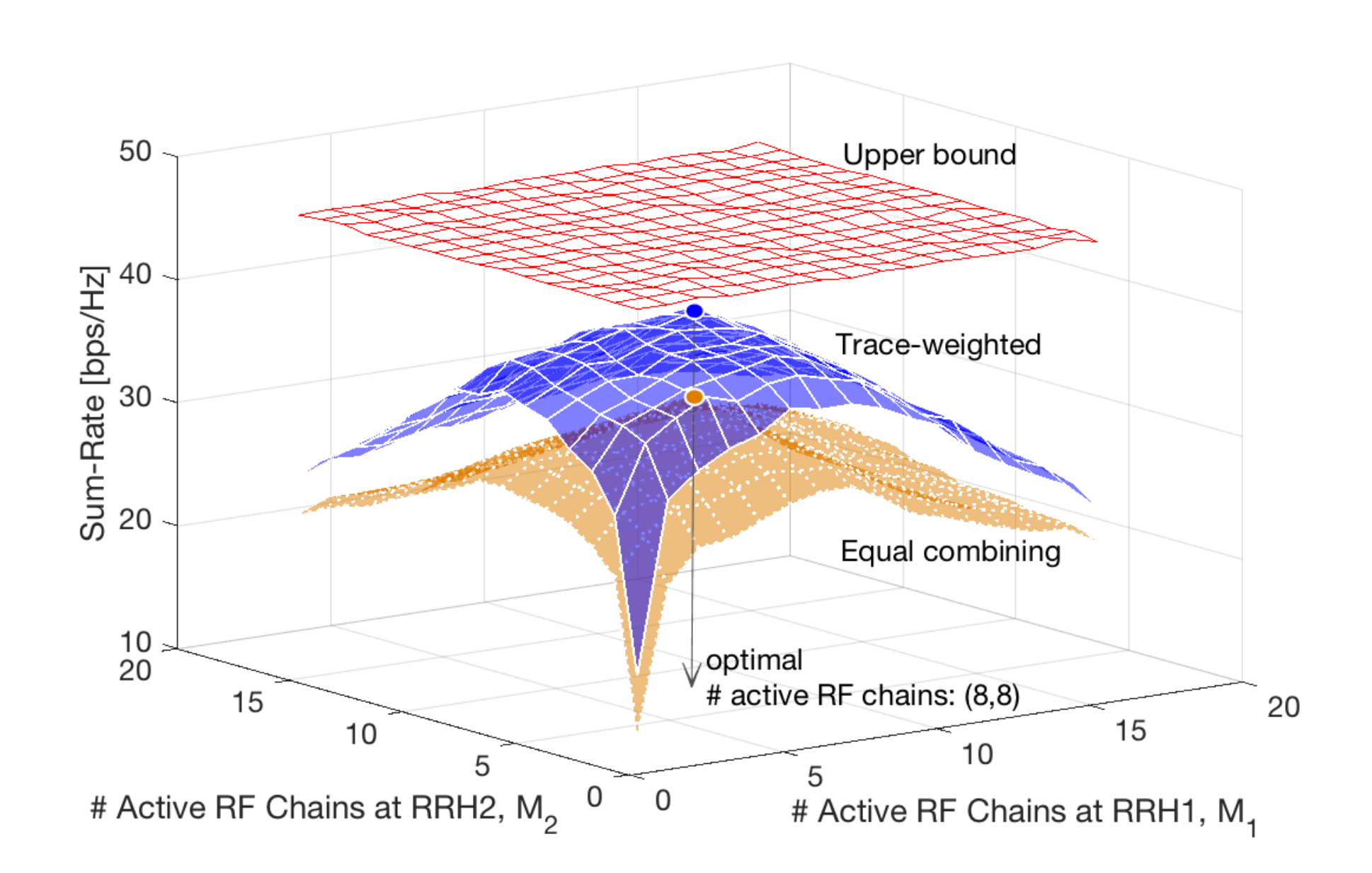}
    \caption{Sum-rate with respect to different $M_1$ and $M_2$, where the maximum sum-rate is achieved at $(M_1^*,M_2^*)=(8,8)$ for both trace-weighted and equal combining methods ($N=64$, $K=3$, $L=2$, $\hat{M} = 16$, $\cf~=~200$~bps, $d_{1,1}=d_{1,2}=1000$~m, $d_{2,1}=d_{2,2}=500$~m, $d_{3,1}=d_{3,2}=100$~m).}
    \label{F:sumrate_rzf_vs_diffrfc}
\end{figure}

To finish, we justify our choice of restricting $M_l = M$\;~$\forall{l}$ via the results presented in Fig.~\ref{F:sumrate_rzf_vs_diffrfc}, where the sum-rate obtained by our proposed designs is evaluated against all possible combinations for active RF chains in a setup with two RRHs ($L=2$). It is observed that, among all possible combinations of $M_1$ and $M_2$, the setting $M_1=M_2=8$ provides the highest sum-rate.

\section{Conclusion}
C-RAN architecture is envisaged to enable distributed massive MIMO systems. The key challenge to enjoy its benefit is the capacity limited fronthaul links between RRHs and the BBU. In this paper, we sought a solution to this problem by adjusting not only the fronthaul compression levels but also the RF chain activations of RRHs. To achieve this goal, we proposed a hybrid precoding design involving a novel analog beamformer that is optimized based on a trace-weighted combining of spatial covariance matrices. The sizes of the analog and digital beamformer, i.e. RF chain activations, are also optimized in order to maximize the large-scale approximated sum-rate. The digital beamformer is accordingly optimized based on instantaneous effective CSI.

We validated the proposed hybrid precoding design by Monte Carlo simulation. The results highlight that exploiting all the RF chains is not always preferable due to the limited fronthaul capacity, which can be optimized via the proposed hybrid precoding algorithm. We also observed the effectiveness of the proposed analog beamformer design based on trace-weighted combining of spatial covariance matrices, which outperforms the existing equal combining approach especially for different path loss channels of UEs, as well as under severely limited fronthaul capacity and/or a small number of total RF chains.

\section*{Appendix -- Proof of Proposition~1}
The proof consists of deriving the deterministic equivalents of the following terms for $\bar{N}\rightarrow \infty$.

\subsubsection{Deterministic Equivalent of $\Psi_{1,l}$}
Applying matrix inversion lemma \cite[Lemma~1]{Wagner:12} twice for $\bh_k^\H \bF_\rf \bC^{-1}$ and $(\bh_k^\H \bF_\rf \bC^{-1})^\H$ in \eqref{Eq:Psi1},
\begin{align}
\Psi_{1,l} &= \sum_{k=1}^K p_k \frac{\bh_k^\H \bF_\rf \bC_{[k]}^{-1} \bB_{\rf,l} \bC_{[k]}^{-1}\bF_\rf^\H \bh_k }{\(1 + \bh_k^\H \bF_\rf \bC_{[k]}^{-1}\bF_\rf^\H \bh_k  \)^2}.
\end{align}

Applying trace convergence lemma \cite[Lemma~4]{Wagner:12} and rank-1 perturbation lemma \cite[Lemma~6]{Wagner:12} yields
\begin{align}
\Psi_{1,l} &\overset{\bar{N}\rightarrow \infty}{\rightarrow} \sum_{k=1}^K p_k \frac{\tr\(\hat{\bR}_k \bC^{-1} \bB_{\rf,l} \bC^{-1} \)  }{\(1 + \bh_k^\H \bF_\rf \bC^{-1}\bF_\rf^\H \bh_k  \)^2}\\
&\rightarrow \sum_{k=1}^K p_k \frac{\tr\(\hat{\bR}_k \bT_{\rf,l}'\)}{\[1 + \tr\(\hat{\bR}_k \bT \)\]^2}
\end{align}
where the last step follows from applying \cite[Theorem~2]{HoydisDebbah:13} to the numerator and \cite[Theorem~1]{Wagner:12} to the denominator.

\subsubsection{Deterministic Equivalents of $\tr(\hat{\bQ}_l)$ and $\Psi_{2,k}$}
By the definitions in \eqref{Eq:Ql},
\begin{align}
\tr(\hat{\bQ}_l)=\sum_{m=1}^{M_l} \tau_{l,m}^2. \label{Eq:PfQl}
\end{align}
According to \eqref{Eq:tau} and \eqref{Eq:fbbkl}, we obtain $\tau_{l,m}^2$ as:
\begin{align}
\tau_{l,m}^2 &=\sum_{k=1}^K p_k \bh_k^\H \bF_\rf \bC^{-1} \bB_{m,l} \bC^{-1} \bF_\rf^\H \bh_k\\
&\overset{(a)}{=} \sum_{k=1}^K p_k \frac{\bh_k^\H \bF_\rf \bC_{[k]}^{-1} \bB_{m,l} \bC_{[k]}^{-1}\bF_\rf^\H \bh_k}{\(1 + \bh_k^\H \bF_\rf \bC_{[k]}^{-1}\bF_\rf^\H \bh_k  \)^2}\\
&\overset{(b)}{\rightarrow} \sum_{k=1}^K p_k \frac{\tr\(\hat{\bR}_k \bC^{-1}\bB_{m,l} \bC^{-1} \)}{\[1 + \tr\(\hat{\bR}_k \bC^{-1}\)\]^2}\\
&\rightarrow \sum_{k=1}^K p_k \frac{\tr\(\hat{\bR}_k \bT_{m,l}' \)}{\[1 + \tr\(\hat{\bR}_k \bT\)\]^2}
\end{align}
where $(a)$ comes from matrix inversion lemma \cite[Lemma~1]{Wagner:12}, $(b)$ from using trace convergence lemma \cite[Lemma~4]{Wagner:12} and rank-1 perturbation lemma \cite[Lemma~6]{Wagner:12}, and the last step is because of applying \cite[Theorem~2]{HoydisDebbah:13} to the numerator and \cite[Theorem~1]{Wagner:12} to the denominator. Applying this to \eqref{Eq:PfQl} leads to the desired result for $\tr(\hat{\bQ}_l)$. As for $\Psi_{2,k}$, utilizing the above result with applying trace convergence lemma \cite[Lemma~4]{Wagner:12} to \eqref{Eq:Psi2} provides the desired result.

\subsubsection{Deterministic Equivalent of $\Psi_{3,k}$} Applying matrix inversion lemma, we obtain the following expression:

\begin{align}
&\Psi_3 = \frac{\bh_k^\H \bF_\rf \bC_{[k]}^{-1} \bF_\rf^\H \bH_{[k]}\bP_{(k)}\bH_{[k]}^\H \bF_\rf \bC_{[k]}^{-1}\bF_\rf^\H \bh_k}{\(1 + \bh_k^\H \bF_\rf \bC_{[k]}^{-1}\bF_\rf^\H \bh_k  \)^2}\\
&\rightarrow \frac{\tr\(\bP_{(k)}\bH_{[k]}^\H \bF_\rf \bC^{-1} \hat{\bR}_k \bC^{-1} \bF_\rf^\H \bH_{[k]} \)}{\[1 + \tr\(\hat{\bR}_k\bC^{-1}\)\]^2} \label{Eq:PfPsi3}
\end{align}
where the last step follows from applying trace convergence lemma \cite[Lemma~4]{Wagner:12} and rank-1 perturbation lemma \cite[Lemma~6]{Wagner:12}. The numerator is rephrased as:
\begin{align}
&\tr\(\bP_{(k)}\bH_{[k]}^\H \bF_\rf \bC^{-1} \hat{\bR}_k \bC^{-1} \bF_\rf^\H \bH_{[k]} \)\\
&= \sum_{i\neq k}^K p_i \bh_i^\H \bF_\rf \bC^{-1} \hat{\bR}_k \bC^{-1} \bF_\rf^\H \bh_i\\
&\overset{(c)}{=}\sum_{i\neq k}^K p_i \frac{\bh_i^\H \bF_\rf \bC_{[i]}^{-1} \hat{\bR}_k \bC_{[i]}^{-1} \bF_\rf^\H \bh_i}{\(1 + \bh_i^\H \bF_\rf \bC_{[i]}^{-1}\hat{\bR}_k \bC_{[k]}^{-1}\bF_\rf^\H \bh_i\)^2} \\
&\overset{(d)}{\rightarrow} \sum_{i\neq k}^K p_i \frac{\tr\(\hat{\bR}_i \bC^{-1}\hat{\bR}_k \bC^{-1}\)}{\[1 + \tr\(\hat{\bR}_i \bC^{-1}\)\]^2} \\
&{\rightarrow} \sum_{i\neq k}^K p_i \frac{\tr\(\hat{\bR}_i \bT_{\hat{\bR}_k}'\)}{\[1 + \tr\(\hat{\bR}_i \bT\)\]^2}
\end{align}
where $(c)$ follows from applying matrix inversion lemma \cite[Lemma~1]{Wagner:12}, $(d)$ from trace convergence lemma \cite[Lemma~4]{Wagner:12} and rank-1 perturbation lemma \cite[Lemma~6]{Wagner:12}, and the last step from utilizing \cite[Theorem~2]{HoydisDebbah:13} for the numerator and \cite[Theorem~1]{Wagner:12} for the denominator. Combining these results with \eqref{Eq:PfPsi3} yields the desired result.

\subsubsection{Deterministic Equivalent of $\Psi_{4,k}$} Applying matrix inversion lemma, we obtain:
\begin{align}
\Psi_4 &= \frac{\bh_k^\H \bF_\rf \bC_{[k]}^{-1}\bF_\rf^\H \bh_k }{1 + \bh_k^\H \bF_\rf \bC_{[k]}^{-1}\bF_\rf^\H \bh_k} \\
&\overset{(e)}{\rightarrow} \frac{\tr\(\hat{\bR}_k \bC^{-1}\)}{1 + \tr\(\hat{\bR}_k \bC^{-1} \)}\\
&\overset{(f)}{\rightarrow} \frac{\tr\(\hat{\bR}_k \bT\)}{1 + \tr\(\hat{\bR}_k \bT \)}
\end{align}\normalsize
where $(e)$ comes from applying trace convergence lemma \cite[Lemma~4]{Wagner:12} and rank-1 perturbation lemma \cite[Lemma~6]{Wagner:12}, and (f) from applying \cite[Theorem~1]{Wagner:12} to the numerator and \cite[Theorem~2]{HoydisDebbah:13} to the denominator. Plugging the aforementioned results into the objective function in \textsf{P1} completes the proof.
\endproof

\bibliographystyle{ieeetr}  
\bibliography{IEEEabrv}

\end{document}